\documentclass[twocolumn,showpacs,preprintnumbers,amsmath,amssymb,superscriptaddress]{revtex4}

\usepackage{graphicx}
\usepackage{dcolumn}
\usepackage{bm}
\usepackage{xr} 
\usepackage{amsmath}
\usepackage{upgreek}
\usepackage{xcolor}
\usepackage[utf8]{inputenc}
\usepackage[T1]{fontenc}
\usepackage{pdfsync}

\begin{document}

\title{Principles for sensitive and robust biomolecular interaction analysis - The limits of detection and resolution of diffraction-limited focal molography}

\author{Andreas Frutiger}
\altaffiliation{%
	A. Frutiger and Y. Blickenstorfer contributed equally to this work.
}%
\affiliation{%
	Laboratory of Biosensors and Bioelectronics, Institute of Biomedical Engineering, ETH Zürich, 8092 Zürich, Switzerland
}%
\author{Yves Blickenstorfer}
\altaffiliation{%
	A. Frutiger and Y. Blickenstorfer contributed equally to this work.
}%
\affiliation{%
	Laboratory of Biosensors and Bioelectronics, Institute of Biomedical Engineering, ETH Zürich, 8092 Zürich, Switzerland
}%
\author{Silvio Bischof}\textbf{}
\affiliation{%
	Laboratory of Biosensors and Bioelectronics, Institute of Biomedical Engineering, ETH Zürich, 8092 Zürich, Switzerland
}%
\author{Csaba Forró}
\affiliation{%
	Laboratory of Biosensors and Bioelectronics, Institute of Biomedical Engineering, ETH Zürich, 8092 Zürich, Switzerland
}%
\author{Matthias Lauer}
\affiliation{
	Roche Pharma Research and Early Development, Roche Innovation Center Basel, 4070 Basel, Switzerland
}%
\author{Volker Gatterdam}
\affiliation{%
	Laboratory of Biosensors and Bioelectronics, Institute of Biomedical Engineering, ETH Zürich, 8092 Zürich, Switzerland
}%

\author{Christof Fattinger}
\email{christof.fattinger@roche.com}
\affiliation{
	Roche Pharma Research and Early Development, Roche Innovation Center Basel, 4070 Basel, Switzerland
}%

\author{János Vörös}
\email{janos.voros@biomed.ee.ethz.ch}
\affiliation{%
	Laboratory of Biosensors and Bioelectronics, Institute of Biomedical Engineering, ETH Zürich, 8092 Zürich, Switzerland
}%

\date{\today}

\begin{abstract}

Label-free biosensors enable the monitoring of biomolecular interactions in real-time, which is key to the analysis of the binding characteristics of biomolecules. While refractometric optical biosensors such as SPR [Surface Plasmon Resonance] are sensitive and well-established, they are susceptible to any change of the refractive index in the sensing volume caused by minute variations in composition of the sample buffer, temperature drifts and most importantly nonspecific binding to the sensor surface. Refractometric biosensors require reference channels as well as temperature stabilisation and their applicability in complex fluids such as blood is limited by nonspecific bindings. Focal molography does not measure the refractive index of the entire sensing volume but detects the diffracted light from a coherent assembly of analyte molecules. Thus, it does not suffer from the limitations of refractometric sensors since they stem from non-coherent processes and therefore do not add to the coherent molographic signal. The coherent assembly is generated by selective binding of the analyte molecules to a synthetic binding pattern – the mologram. Focal Molography has been introduced theoretically [C. Fattinger, Phys. Rev. X 4, 031024 (2014)] and verified experimentally [V. Gatterdam, A. Frutiger, K.-P. Stengele, D. Heindl, T. Lübbers, J. Vörös, and C. Fattinger, Nat. Nanotechnol. 12, 1089 (2017)] in previous papers. However, further understanding of the underlying physics and a diffraction-limited readout is needed to unveil its full potential. This paper introduces refined theoretical models which can accurately quantify the amount of biological matter bound to the mologram from the diffracted intensity. In addition, it presents measurements of diffraction-limited molographic foci i.e. Airy discs. These improvements enabled us to demonstrate a resolution in real-time binding experiments comparable to the best SPR sensors, without the need of temperature stabilisation or drift correction and to detect low molecular weight compounds label-free in an endpoint format. The presented experiments exemplify the robustness and sensitivity of the diffractometric sensor principle. 

\end{abstract}


\maketitle


\section{Introduction}

\begin{figure*}[t]
	\includegraphics{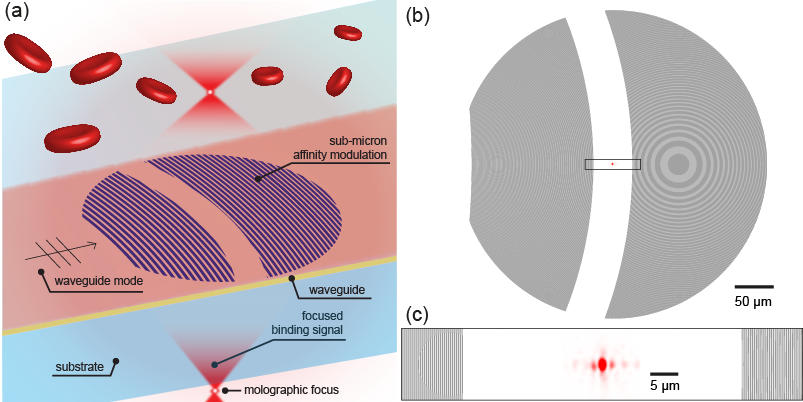}
	\caption{\label{fig0} Focal molography incorporates the four essentials of a highly sensitive diffractometric biosensor (a) A sub-micrometer affinity modulation of specific binders is exposed to a biological sample [i.e. blood]. The mode of a high refractive index waveguide provides perfect darkfield illumination of the molecules in the vicinity of the sensor surface and enhances the light intensity. The shape of the pattern acts as a diffractive lens, which concentrates the diffracted signal into a focal spot, whereas the background intensity is diluted over the entire solid angle. (b) The molographic pattern with the actual focal spot superimposed [bottom view] and enlarged in (c). The Airy disk of the mologram [red dot] monitors the binding activity of billions of recognition sites on an area that is nearly five orders of magnitude larger than the tiny focal spot.}
	
\end{figure*}

Diffractive lenses or focusing holograms proposed by Augustin-Jean Fresnel are known to humanity since two hundred years and have ever since experienced applicability in various fields such as photography \cite{Mohammad2018-bb}, telescopes \cite{Early2004-gz}, spectroscopy \cite{Park2008-yn}, optical tweezers \cite{Rodrigues_Ribeiro2017-jj}, and X-ray lenses \cite{Mohacsi2014-ol}. Yet, nature has discovered this principle much earlier. In certain organisms, biomolecules are assembled to create a focusing hologram for image formation in the eye \cite{Palmer2017-cl}. Recently, thanks to advances in photo-lithography \cite{Gatterdam2017-oo} and non-fouling photoactivatable surface chemistries in particular \cite{Serrano2016-hy}, it became possible to apply the holographic principle to highly sensitive molecular detection. These molecular holograms can be used for real-time label-free detection of molecules by molecular recognitions in complex samples \cite{Gatterdam2017-oo}. This enables the monitoring of biomolecular interactions, which is key to the analysis of binding characteristics of biomolecules in a broad range of applications \cite{Fraser2017-vs}. To date, the biosensing field is dominated by refractometric optical sensors and most prominently, thanks to their high surface sensitivity, by techniques based on evanescent waves, such as surface plasmons or dieletric waveguide modes \cite{Homola2006-uz,Kozma2014-tv}. These analytical tools are well-established to perform label-free binding assays with high sensitivity and low limits of detection [0.1-1 pg/mm$^2$] \cite{Cannon2004-kf,Kozma2009-tr}. Refractometric biosensors [e.g. surface plasmon resonance, SPR] measure the refractive index change upon receptor-ligand binding in the vicinity of the sensor surface. However, they are susceptible to any change in refractive index within the evanescent field caused by fluctuations in temperature, buffer composition and most importantly nonspecific binding to the sensor surface. This inherent feature of refractometric sensors often manifest as drift and causes jumps in the sensor signal - e.g. during sample exchange \cite{Piliarik2010-fb}. Therefore, these sensors typically operate continuously and in well-defined buffers because measurements in serum or plasma exhibit artifacts and stability problems.

The mentioned limitations arise from the fundamental inability of a refractometric sensor to distinguish between the molecules of the target analyte and all other influences that affect the refractive index of the sensing volume. Even for evanescent field sensors, the sensing volume is still enormous compared to the small volume of the target molecules. This makes it virtually impossible to compensate for these influences - even with a differential measurement \cite{Piliarik2010-fb}. There is, however, a physical phenomenon that only measures the refractive index difference between the target molecules and the refractive index of their displacement volume, namely the scattering of light. This rejects most of the influences from temperature and buffer changes by measuring only the refractive index contrast in the nanoenvironment of the binding events.

It is a common belief in the biosensing community that the most sensitive detection methods for label-free biomolecular interaction analysis in real-time are based on the refractometric sensing principle. In this context it is sometimes not esteemed that biomolecular interactions can be detected with high sensitivity by the scattering of light \cite{Piliarik2014-ys,Fattinger2014-dm,Gatterdam2017-oo}. The single molecule detection method iSCAT \cite{Piliarik2014-ys} is based on interferometric detection of scattering. It demonstrates exquisite sensitivity for the analysis of biomolecular interactions through scattering. 
The single molecular sensitivity of iSCAT allows analyzing the heterogeneity in an ensemble of a molecular species. Yet, in other applications it is sufficient to determine an averaged quantity of the ensemble. The accurate quantification of a biomarker concentration falls in this category. In such a measurement, single molecular sensitivity can be beneficial but is neither required nor should its importance be overestimated. In the case of iSCAT, the single molecular sensitivity comes at the cost of a relatively complicated setup, since the noise has to be sufficiently low to detect every single protein independently. In addition, iSCAT cannot distinguish between different types of similarly sized proteins. Therefore, for measuring in complex fluids, iSCAT is currently limited by non-specific binding similarly to refractometric sensors. In both cases, the specificity is mostly determined by the choice of surface chemistry \cite{Piliarik2014-ys}. A protein-repellent [non-fouling] surface chemistry is not enough to measure in complex samples since there is always a significant amount of defects in the ad-layer and therefore of nonspecific binding to the sensor \cite{Pasche2003-mr,Love2005-oq}.

\newpage

Conversely, molecular holograms are diffractometric sensors, which offer an additional mechanism to reduce the effect of nonspecific binding. This is achieved by constraining the specific binding to a coherent scattering system - i.e. a molecular hologram. The blueprint of this hologram is encoded into the surface ad-layer. Namely, the recognition sites compose a coherent binding pattern - i.e. a mologram. The constructive interference of the scattered fields relates all bound analyte molecules and yields a quadratic scaling of the measured intensity with respect to the analyte number. On the other hand, the scattered field of randomly bound background molecules interferes with equal probability constructively or destructively. Therefore, only its variance affects the coherent signal and thus it scales linearly with particle number. This implies that the nonspecific binding is suppressed efficiently for a sufficiently large ensemble of analyte molecules. In addition, other random scattering [noise] sources experience the same repression with respect to the signal. Therefore, it is considerably simpler to detect an ensemble of molecules coherently than to count them individually. In other words, a diffractometric sensor is inherently self-referencing on the sub-micron length scale of regions of constructive and destructive interference. 

A pure diffractometric biosensor consists of coherently arranged binding sites without any diffractive power, or in other words a massless affinity modulation \cite{,Fattinger2014-dm}. A sensor with these properties is extremely robust and only produces a signal in the presence of the analyte \cite{Gatterdam2017-oo}. The conception of an affinity modulation that has no optical modulation is reasonable to physicists. However, advanced molecular engineering capabilities are required to achieve this experimentally, since most nanolithographic techniques such as imprinting or lift-off techniques produce an inherent optical modulation due to coherent defects in the affinity keys \cite{,Falconnet2004-mq,Vigneswaran2014-ka}. Even worse, these coherent defects will give rise to an affinity modulation for background molecules, severely compromising the rejection of nonspecific binding. This fact makes the implementation of sensitive diffractometric biosensors interdisciplinary and demanding.

 Focal molography is the first diffractometric sensor that may exhibit resolutions in direct binding assays comparable to the best refractometric sensors \cite{Gatterdam2017-oo,Fattinger2014-dm}. Briefly, in focal molography the mologram is situated on a high refractive index slab waveguide and illuminated by the fundamental TE mode [Fig. \ref{fig0}]. When the affinity modulation is exposed to a biological sample the analyte binds to the mologram. This induces an optical grating that diffracts light from the guided TE mode into a diffraction-limited focal spot. From the diffraction efficiency, the presence of molecules at the interaction sites is quantified. We will now outline the four pillars that need to be fulfilled for diffractometric biosensors to be highly sensitive and robust. Previously reported diffractometric concepts for biomolecular interaction analysis  \cite{Avella-Oliver2017-jb,Cleverley2010-qm,Lenhert2010-tt,Lai2008-ha}\footnote{An error in Eq. (2)  of Ref. \cite{Lai2008-ha} should be noted: $\Lambda$ does not refer to the grating period but to the wavelength of the n=-1 wave inside the grating region in the direction normal to the waveguide surface as described in Ref. \cite{Tamir1977-xc}.} do not incorporate all four pillars. This is the reason why they are limited in sensitivity or robustness. The concept - focal molography - was introduced with all the essentials necessary for highly sensitive and robust diffractometric biomolecular interaction analysis \cite{,Fattinger2014-dm} [Fig. \ref{fig0}]. [I] The first essential is a sub-micrometer affinity modulation on a non-fouling monolithic surface-layer for efficient rejection of nonspecific binding. In the first demonstration of focal molography, this has been achieved by the reactive immersion lithography [RIL] process, which produces an affinity modulation consisting of active regions [ridges] and passive regions [grooves] on a non-fouling brushed copolymer ad-layer \cite{,Gatterdam2017-oo}. Ideally, this affinity modulation should be massless. [II] In focal molography, the mologram is situated on an asymmetric high refractive index slab waveguide, which provides the second essential - a proper dark-field illumination of the coherent affinity modulation. The two dimensional light sheet of the guided TE mode only illuminates the first 100 nm of the sample solution close to the surface. This avoids any background scattering from particles in the sample solution that are further away. [III] The high refractive index waveguide also provides the third essential, namely an increase in the field intensity at the scatterer location. In other words, guided to free-space mode coupling is more efficient than free-space to free-space coupling for a given amount of coherent biological matter \cite{,Kogelnik1969-pf}. [IV] The fourth and last essential is the observation of the diffracted signal in the far-field of the mologram defined in terms of Fraunhofer distance. This near to far-field transformation increases the signal to noise ratio due to the directed character of the diffracted signal compared to the distributed background from random scatterers. However, the Fraunhofer distance of a linear diffraction grating with a length of 400 $\upmu$m is roughly 50 cm for visible wavelengths. By using a lens, the far-field can be observed much closer to the sensor surface. In our case, the mologram itself performs the near to far-field transformation by focusing the intensity holographically onto an Airy disk only a few hundred microns away from the sensor surface. The binding information of billions of recognition sites on an area five orders of magnitude larger is therefore contained in the tiny Airy disk [Fig. \ref{fig0}(b),(c)]. This enables compact technical realizations of diffractometric biosensors. 
 
 From another viewpoint, focal molography can also be seen as a “chemical radio”, at least in the eyes of a physical chemist \cite{,Poole1998-pk}. The transmission of radio signals is based on the modulation of an RF carrier signal and the subsequent demodulation at the receiver. Molography applies this principle at optical frequencies to the transmission of chemical signals. Molecules recognize the affinity modulation in the mologram and interact with it. The molecular interaction renders a coherent molecular pattern in the form of a diffractive lens. This diffractive lens modulates the momentum of the guided mode with the spatial frequency of the mologram. The demodulation in k-space is performed by Fourier optics and the molographic signal is separated from the carrier wave in the focal plane of the lens. 

Recently, the first experimental measurements with focal molography have been performed using the non-diffraction-limited ZeptoReader [Zeptosens AG], substantially compromising the fourth pillar \cite{,Gatterdam2017-oo}. The emphasis of that publication was the demonstration of the robust operation of focal molography and its insensitivity to nonspecific binding in complex samples rather than achieving high sensitivity. Nevertheless, a real-time detection limit of 5 pg/mm$^2$ was achieved and we made the projection that noise levels can be reduced by at least two orders of magnitude by observing the molographic signal in a reader capable of resolving the diffraction-limited focus.

\begin{figure*}[t]
	\includegraphics{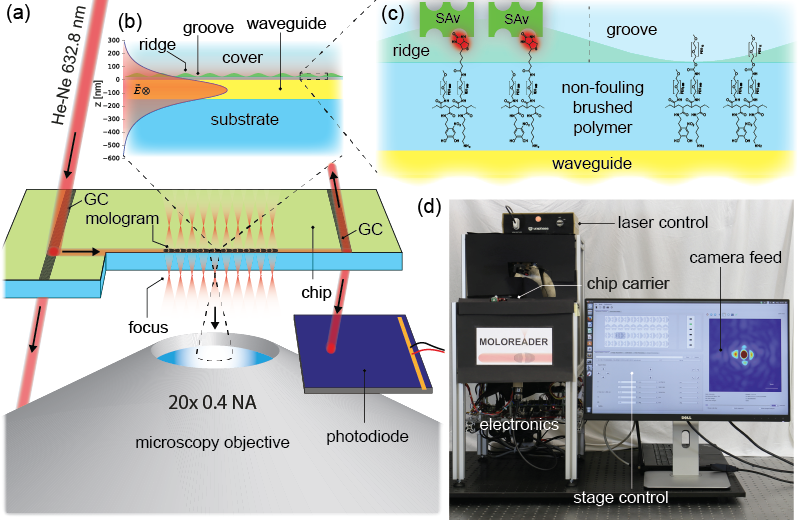}
	\caption{\label{fig1}Detection of diffraction-limited molographic spots (a) Schematic representation of the setup for the experimental demonstration of diffraction-limited molography: Light of a He-Ne laser is coupled into the fundamental TE mode of a high refractive index slab waveguide via a grating coupler [GC]. The light propagates along the waveguide and is scattered at the molecules of interest [Streptavidin] that form a focusing hologram. These molecules are captured from solution by binding to a coherent affinity modulation on top of a non-fouling polymer layer that is fabricated by reactive immersion lithography \cite{Gatterdam2017-oo}. The molographic signal [the intensity of the focal point] of one of ten molograms in a row is collected by a microscope objective and captured by a camera. The total power in the waveguide is monitored with a photodiode, which measures the light diffracted by a physical out-coupling grating which is etched into the waveguide. (b) Enlarged view of the waveguide with the field profile of the TE mode. (c) In active regions [ridges] immobilized receptors [biotin] capture the protein of interest [SAv] and form a coherent assembly, whereas inert regions backfilled with polyethylene glycol [grooves] do not recognize the protein. (d) The experimental setup [MoloReader] in operation. The microscopy objective is focused on the focal plane of the mologram.}
	
\end{figure*}

The aim of this contribution is therefore to characterize the molographic signal in the proper far-field and thus to explore the resolution limits of diffraction-limited focal molography for massless affinity modulations experimentally as well as to refine some of the theoretical concepts. First, we introduce a measurement setup that allows static and real-time measurement of diffraction-limited molographic focal spots. Second, we present a semi-analytical framework with which the field distribution in the focal spot can be accurately computed by summation of the scattered fields of individual molecules [dipole scatterers] on the waveguide surface. Third, we demonstrate that the simulated and experimental field distributions are in excellent agreement with each other and that the Airy disk dimensions of the mologram are consistent with the Airy disk of a diffraction-limited lens. Fourth, we show that our synthetic holograms produce diffraction-limited focal spots at least up to mologram diameters of 400 $\upmu$m on high refractive index slab waveguides. Furthermore, it was verified that the analytical predictions for the intensity of the focal spot through coupled mode theory made by Fattinger \cite{,Fattinger2014-dm} coincide with Rayleigh scattering and can accurately describe the experimentally measured intensities for a given amount of coherent biological matter. Fifth, we address the relevance of different background sources that can scatter intensity into the focal plane and produce an inhomogeneous speckle pattern that limits the resolution and accuracy of the molographic measurement. Based on this discussion, a figure of merit for molography is formulated that allows direct comparison of different molographic arrangements with different waveguides and mologram sizes. Next, we apply our theoretical insights to calculate limits of detection for molography on Ta$_2$O$_5$ slab waveguides for endpoint and real-time detection. These predictions are then verified experimentally. In particular, the low molecular weight [< 300 Dalton] molecule vitamin B7, commonly known as biotin, is detected label-free by molography in an endpoint measurement without any calibration of the sensor. These biotin molograms are most likely the faintest man made holograms that have ever been measured. In addition, we demonstrate that it is possible to fabricate an affinity modulation without a detectable optical modulation and use it to acquire real-time binding curves with 500 pM Streptavidin [SAv] in buffer that exhibit baseline noise levels below 100 fg/mm$^2$ over 20 mins which are comparable to the best commercially available label-free detection method \cite{noauthor_undated-ex}. However, while the commercial system is temperature stabilized to 0.01 $^\circ$C, we achieved this stability without any temperature control demonstrating the potential of focal molography for extremely sensitive and robust, real-time, label-free molecular interaction analysis.


\section{Diffraction-limited molography} 
\label{sec:diffraction_limited_focal_spots_measurement_and_simulation}

\subsection{Measurement of foci formed by diffraction-limited molograms} 
\label{sub:measurement_of_diffraction_limited_molograms}

The realization and quantification of diffraction-limited molographic experiments incorporates the design of a microscope, waveguide coupler, fluidics as well as the development of appropriate algorithms for evaluation of the acquired images. The setup [MoloReader]  which we developed for this purpose  is displayed in Fig. \ref{fig1}(d), as well as in Fig. \ref{fig:MoloreaderConfigureation} and a functional schematic is shown in Fig. \ref{fig1}(a). The setup allows to couple a TE polarized He-Ne laser beam [632.8 nm wavelength] via a grating coupler [coupling angle -10.6$^{\circ}$, period 318 nm, length 500 $\upmu$m] into a thin-film optical waveguide [145 nm thick Ta$_2$O$_5$] on a glass substrate [D263 Schott, 700 $\upmu$m]. Molecules located on the waveguide are illuminated by the evanescent field of the fundamental guided TE mode [$N=1.814$, penetration depth 82 nm] in a dark-field manner [Fig. \ref{fig1}(b)]. For most of the experiments presented in this paper, the molograms composed of binding sites for the protein molecule Streptavidin [SAv]. The molograms on the waveguide consist of alternating \textit{ridges}, where SAv binds to immobilized biotin [MW: 227 g/mol]; and \textit{grooves}, which are backfilled with an inert PEG molecule [MeO-dPEG$_{12}$, MW: 570 g/mol]. The PEG backfilling is performed to obtain a massless affinity modulation, as well as for blocking of free amine groups. As a side note, despite the higher molecular mass of the PEG, these molograms exhibit a non-detectable mass modulation [Movie 3]. Most likely because PEG is the more flexible molecule and has a smaller refractive index increment than biotin [0.12 compared to 0.16 ml/g]. The SAv bound mologram is denoted as [NH-biotin/SAv|NH-PEG$_{12}$] and was fabricated by reactive immersion lithography as introduced previously [Fig. \ref{fig1}(c)] \cite{,Gatterdam2017-oo}. We developed a new version of the illumination setup that achieved higher peak-to-peak mass modulations of 540 pg/mm$^2$ [27 \%] compared to the previously published 283 pg/mm$^2$ [14 \%] \cite{,Gatterdam2017-oo} thanks to the higher spectral and spatial coherence of the laser source used [405 nm] [Figs.  \ref{fig:Setup} and \ref{fig:CharacterisationSetup}]. This value was determined from a STED [stimulated emission depletion microscopy] measurement on a Leica SP8 STED as described in our previous publication \cite{,Gatterdam2017-oo}. When impinging on the mologram, a small portion of the light is coupled out into two converging beams that form two diffraction-limited foci above and below the waveguide. The diffracted light of the lower beam is collected by a 20 x, 0.4 NA microscopy objective and visualized by a CMOS camera. The molographic pattern has a diameter of 400 $\upmu$m, a numerical aperture of 0.33, a focal length of 900 $\upmu$m in glass [$n_{\rm{s}} = 1.521$] and a sickle shaped central recess area [Bragg recess area] of 50 $\upmu$m width to avoid second order Bragg reflections \cite{,Fattinger2014-dm}. The recess area is formed by two concentric circles with 1040 $\upmu$m and 1140 $\upmu$m diameter, respectively. 


\subsection{From protein molecules to molographic signals - Simulations of molographic foci} 
\label{sub:numerical_simulations}

\begin{figure}
	\includegraphics{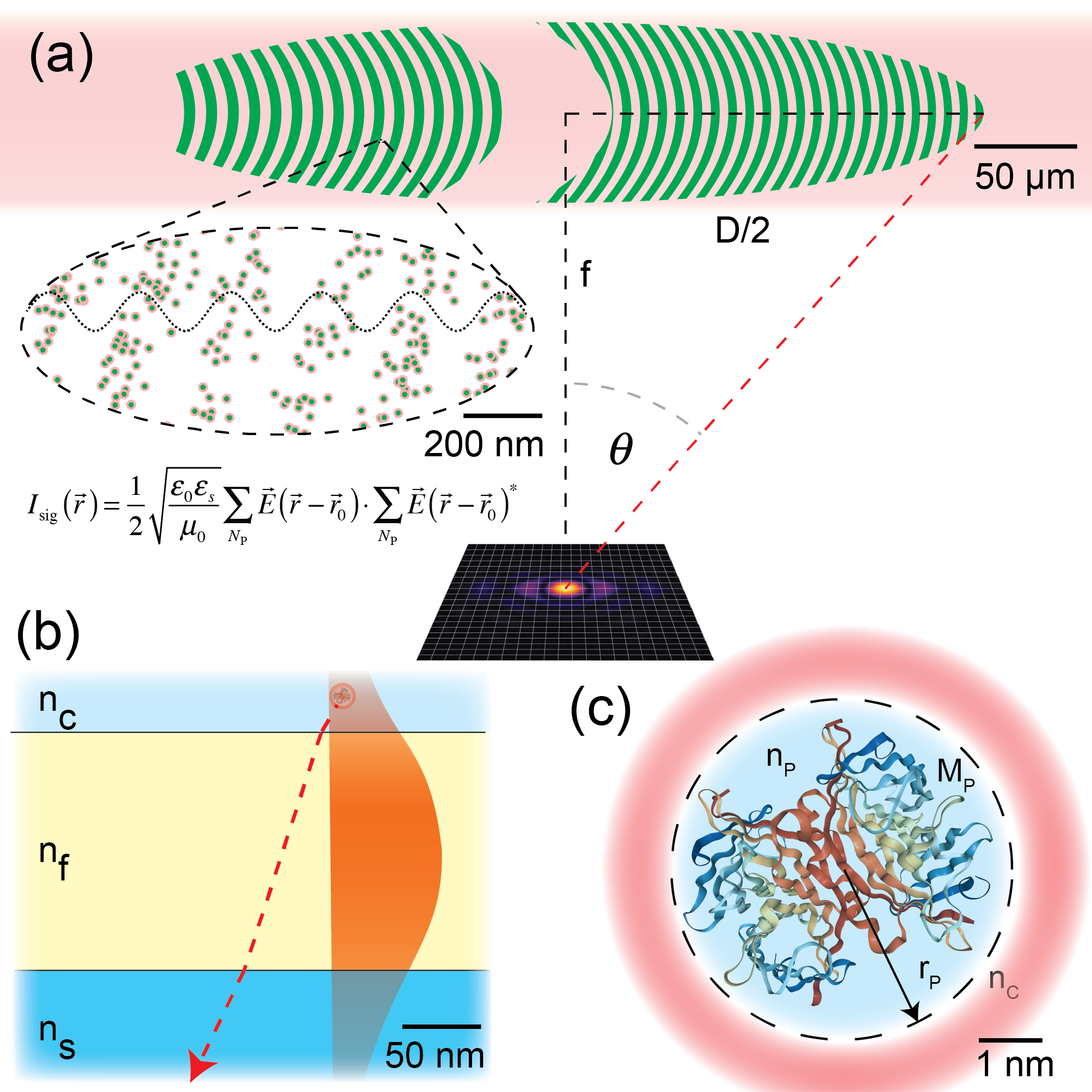}
	\caption{\label{fig2}Simulation of molographic foci (a) The molographic signal emerges from the superposition of the scattered electric fields of many individual protein molecules on the surface of the waveguide [proteins are not drawn to scale but their number density corresponds to the 2.6 pg/mm$^2$ at the detection limit [Fig. \ref{fig7}]]. This field is computed for every pixel on a specified screen in the focal plane of the mologram. (b) The scattered field is calculated by modeling the proteins as Rayleigh scatterers excited by the evanescent field of the waveguide mode, which is obtained by solving the eigenvalue problem of the slab waveguide \cite{Marcuse1974-uo}. $n_{\rm{c}}$, $n_{\rm{f}}$ and $n_{\rm{s}}$ are the refractive indices of cover, film and substrate. The dielectric interfaces can be accounted for by computing the dipole potentials of the two layer interface as described by Novotny and Hecht \cite{Novotny2011-pa}. (c) The optical properties necessary to determine the polarizability of the protein dipole i.e. refractive index and radius can be calculated from its molecular mass and the refractive index increment for proteins in water \cite{Fattinger2014-dm}. } 
\end{figure}

The qualitative intensity distribution in and around the diffraction-limited molographic spot can be described by the coherent superposition of individual Rayleigh scatterers or by a mean-field approach through Fourier optics both yielding the same results \cite{,Goodman2005-wy}. Here we chose the first method to investigate the expected intensity distribution in the focal plane by summation of the scattered electric field of individual dipoles [molecules] located on the mologram [Fig. \ref{fig2}(a)]. We wrote a GPU-based Python framework that can simulate the intensity distribution of a large amount [few 100 millions] of scatterers on a plane with typically 150 x 150 pixels resolution within roughly an hour. This semi-analytical approach has the great advantage of calculating the field only where it is to be determined [compared to FDTD or FEM approaches]. The exact procedure is outlined in the Supplemental Material [SM] Section \ref{sec:description_of_the_simulation_framework} and shall only be summarized here. First, the eigenmode equation of the dielectric waveguide was solved according to Marcuse \cite{,Marcuse1974-uo} for the fundamental TE mode in order to calculate the excitation field at the position of the scatterers, which were placed on the molographic pattern [Fig. \ref{fig2}(b)]. Proteins can be modeled as Rayleigh scatterers due to their small size of only a few nm \cite{,Sinclair1947-xa}. The dipole strength of a protein molecule depends only on its molecular mass and the immersion medium [Fig. \ref{fig2}(c)]. This is because the radius and the refractive index of the resulting sphere are related to the molecular mass and can be calculated as described in the SM Section \ref{sub:model_of_scattering_protein_particles_in_water}. For most practical purposes the exact composition of the protein is negligible for its scattering properties. To account for reflections at the optical interfaces, we used the dipole potential approach outlined by Novotny and Hecht \cite{Novotny2011-pa}. Furthermore, multi-body interactions were disregarded because the scattering cross-section of a typical protein is only in the order of 10$^{-24}$ m$^2$ and the total diffracted power is typically less than 1 \%.


\subsection{Comparison between analytical, numerical and experimental results} 
\label{sub:comparison_analytical_numerical_experimental}

\subsubsection{Shape of the molographic Airy disk} 
\label{ssub:shape_of_the_airy_disk}

\begin{figure}
\includegraphics{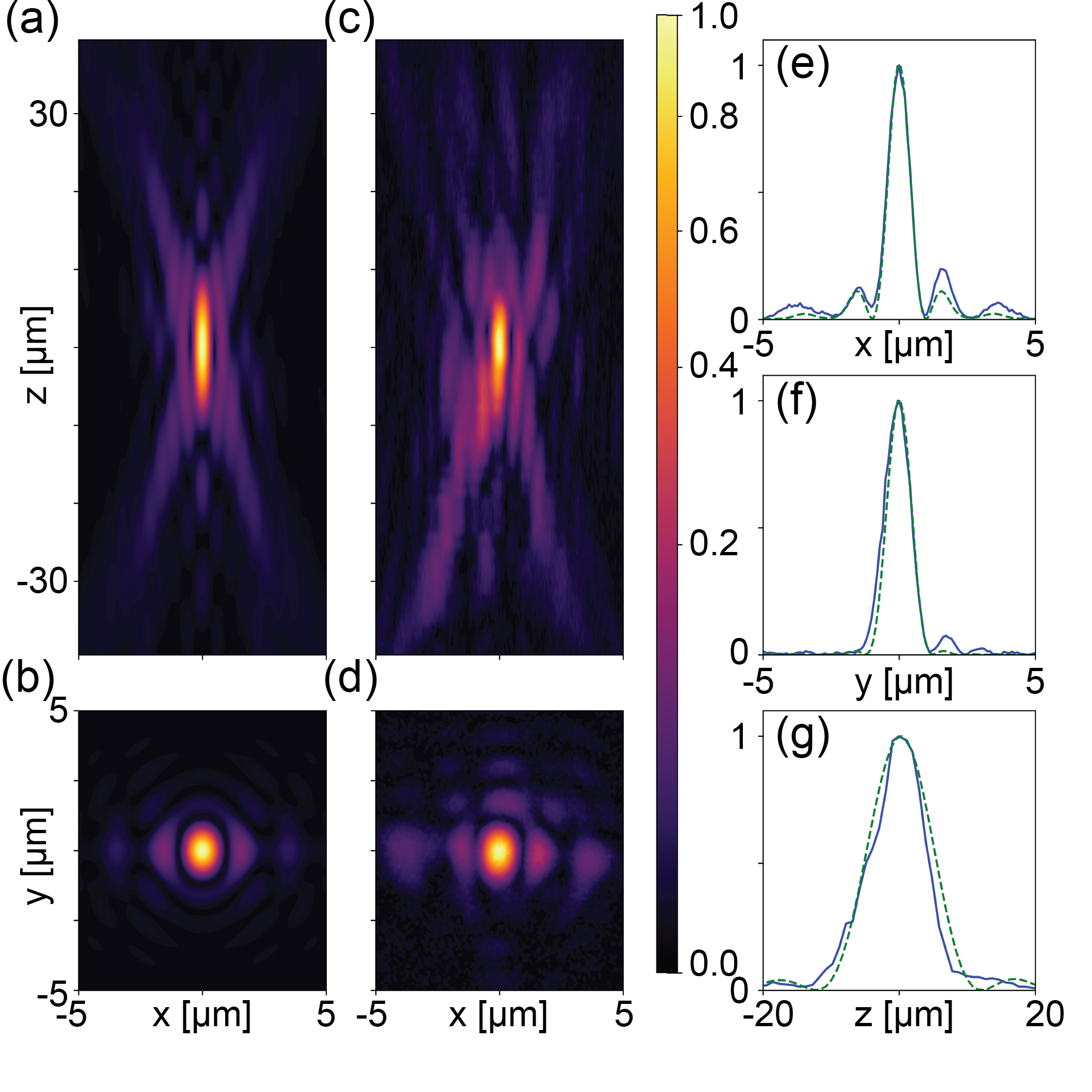}
\caption{\label{fig3}The normalized intensity distribution in the vicinity of the focal spot. The contour plots show the vertical and the horizontal focal plane of the normalized intensity signal obtained by simulations (a),(b) and experiments (c),(d). The line plots (e)-(g) show the cross sections evaluated through the focal point along each axis. Both the simulated [dashed green] and the experimentally obtained [solid blue] curves in the focal plane exhibit the shape of a Bessel function as expected from ideal lenses. The Airy disk is slightly enlarged in y-direction due to the sickle shaped recess area in the middle of the mologram \cite{Fattinger2014-dm}.} 
\end{figure}

Fig.  \ref{fig3} illustrates the diffraction-limited focus obtained by simulations [Figs. \ref{fig3}(a),(b)] and experiments [Figs. \ref{fig3}(c),(d)] as an axial and radial slice through the focal point of the mologram. The experimental molographic spot was acquired from a SAv$^{555}$ [Alexa Fluor$^{\rm{TM}}$ 555 labeled, Thermo Fisher Scientific] mologram in water. The fluorophore was only used for quality control and its scattering cross section is negligible compared to the one of SAv. Therefore, we will only refer to SAv for the rest of the paper. The chip was fabricated by reactive immersion lithography in DMSO \cite{,Gatterdam2017-oo} followed by a 15 min incubation of 1 $\upmu$M SAv in PBS-T buffer [pH 7.4; 0.05 \% Tween20]. This yields a 540 pg/mm$^2$ peak-to-peak sinusoidal surface mass modulation. The experimental focal spot was acquired under water immersion by pipetting 10 $\upmu$l of DI water on the chip and performing a z-stack with the MoloReader [vertical resolution 1.3 $\upmu$m]. The computed focal spot was obtained from a simulation of 4.75 million SAv molecules [7.9 pg/mm$^2$ peak-to-peak modulation] sinusoidally distributed on the ridges of the mologram with water as the cover medium. This amount of SAv molecules was sufficient to demonstrate the excellent agreement of the numerical results with the measured experimental intensities. The simulated and experimental images only differ by the speckle pattern caused by scattering of the guided wave at non-coherent dipoles, which were not taken into account in the simulations. 

The Airy disk radius for a diffraction-limited lens is determined by the wavelength and the numerical aperture of the mologram $\Delta x = 0.61\frac{\lambda }{{\rm{NA}}}$ which leads to 1.17 $\upmu$m for our molograms. The Airy disk radii found in the measured [solid blue] and the simulated curves [dashed green] in Figs. \ref{fig3}(e) and \ref{fig3}(f), are 1.09 $\upmu$m and 1.07 $\upmu$m in x and 1.34 $\upmu$m and 1.22 $\upmu$m in y-direction, respectively. The Airy disk is slightly elongated in y-direction in both experiment and simulation due to symmetry breaking of the central Bragg recess area of the mologram [Fig. \ref{fig2}(a)]. Without considering the central recess area the Airy disk is perfectly symmetrical and has the size of the focal spot of a diffraction-limited lens [Fig. \ref{fig:No_Bragg}]. In the experiment, there is additional broadening by scattering of the guided mode at waveguide imperfections into other guided modes with a small y-component in the propagation vector. In the extreme case of a contaminated [strongly scattering] waveguide, the molographic spot attains a sickle shape [so called m-line \cite{,Monneret2000-im}]. The depth of field of the mologram also follows the equation for a diffraction-limited lens: $\Delta z = 2{n_{\text{s}}}\frac{\lambda }{{{\text{N}}{{\text{A}}^2}}}$ [Chapter 4 in Ref. \cite{,Novotny2011-pa}]. The depth of field depends on the refractive index of the medium in which it is observed. Since the thickness of our chip is smaller than the focal length of the mologram, we observe the molographic focus in air [Fig. \ref{fig:Effective_Focal_Point}]. Yet, the simulation was carried out in an infinitely thick glass slide, therefore the depth of field had to be compressed by a factor 1/1.521 to match the experiment. After this adjustment, the experimental and the simulated depth of field amounted to 11.23 $\upmu$m and 12.61 $\upmu$m, respectively [Fig. \ref{fig3}(g)]. These are in close agreement with the expected value of 11.62 $\upmu$m for a diffraction-limited lens.

\subsubsection{Quantitative intensity in the molographic focus by analytical predictions, simulations and experiments} 
\label{ssub:quantitative_intensity_in_the_focal_spot_by_analytical_predictions_simulations_and_experiments}

The quantitative intensity within the Airy disk is amenable through coupled mode theory [CMT] or summation of Rayleigh scatterers [RS] without considering multiple reflections of the scattered light at the interfaces of the waveguide. Here, we briefly show that the analytical expressions for the two approaches are equivalent and verify them by numerical simulations and experiments. 

Fattinger \cite{Fattinger2014-dm} used coupled mode theory to obtain an expression for the ratio of power diffracted from the molecular assembly to the power guided by the waveguide [both powers are expressed in power per unit length] \cite{Tamir1977-xc}. Here, we adapt this expression to yield the intuitive transfer function between the intensity on the waveguide surface and the average intensity in the Airy disk [derivation in the SM Section \ref{sec:Molographic_signal_intensity_from_coupled_mode_theory_and_Rayleigh_scattering}] which reads 
\begin{equation}
 	\label{I_sg_CMF}
 	{I_{{\rm{avg,CMT}}}} = 5.59 \cdot {\rm{N}}{{\rm{A}}^2}{\left( {\frac{{dn}}{{dc}}} \right)^2}\frac{{{D^2}}}{{{\lambda ^4}}}{{{\eta _{{\text{mod}}}}{{_{\left[ {\text{A}} \right]}}^2}{\Delta _\Gamma }^2}}{I_0}
 \end{equation}
 The subscript CMT stands for coupled mode theory. $\rm{NA}$ is the numerical aperture of the mologram, ${\frac{{dn}}{{dc}}}$ the refractive index increment for proteins in water \cite{Zhao2011-ju}, $D$ the diameter of the mologram, $\lambda$ the wavelength. Here, we have adapted and generalized the canonical surface mass modulation $\Delta _{{\Gamma _{{\rm{can}}}}}$ introduced by Fattinger \cite{Fattinger2014-dm} with the concept of the analyte efficiency of the modulation ${\eta _{{\text{mod}}}}{{_{\left[ {\text{A}} \right]}}}$. The analyte efficiency of the modulation is analogous to the diffraction efficiency of gratings with different grating functions \cite{Magnusson1978-sx}. The surface mass density modulation can be computed from ${\Delta _\Gamma } = \frac{{{m_{{\text{mod}}}}}}{{{A_ + }}}$, where $m_{\rm{mod}}$ is the mass of the modulation and ${{A_ + }}$ the area of the ridges [see SM Section \ref{sec:definitions}]. For a sinusoidal surface mass density modulation [obtained to a first approximation from phase mask lithography] this is equal to the peak-to-peak value. Here we note that Fattinger defined the canonical surface mass density modulation differently. His definition would correspond to the molographic surface mass density [see SM Section \ref{sec:definitions}] and is therefore a factor two smaller.  The prefactor in Eq. \eqref{I_sg_CMF} arises from various considerations such as taking into account the relative power incident on the Airy disk and its size. The intensity on the waveguide surface is given by

\begin{equation}
\label{intensity_on_wg_surface}
	{I_0} = 2 \frac{{{n_c}\left( {n_{\text{f}}^2 - {N^2}} \right)}}{{N{t_{{\text{eff}}}}\left( {n_{\text{f}}^2 - n_{\text{c}}^2} \right)}}{P_{{\text{wg}}}}
\end{equation}
where $P_{\rm{wg}}$ is the power per unit line [W/m] in the waveguide, $t_{\rm{eff}}$ the effective thickness of the waveguide, $N$ the effective refractive index of the fundamental TE mode, $n_{\rm{f}}$, $n_{\rm{c}}$ the refractive indices of the waveguide film and the cover medium.

\begin{figure}
\centering
\includegraphics[width=0.46\textwidth]{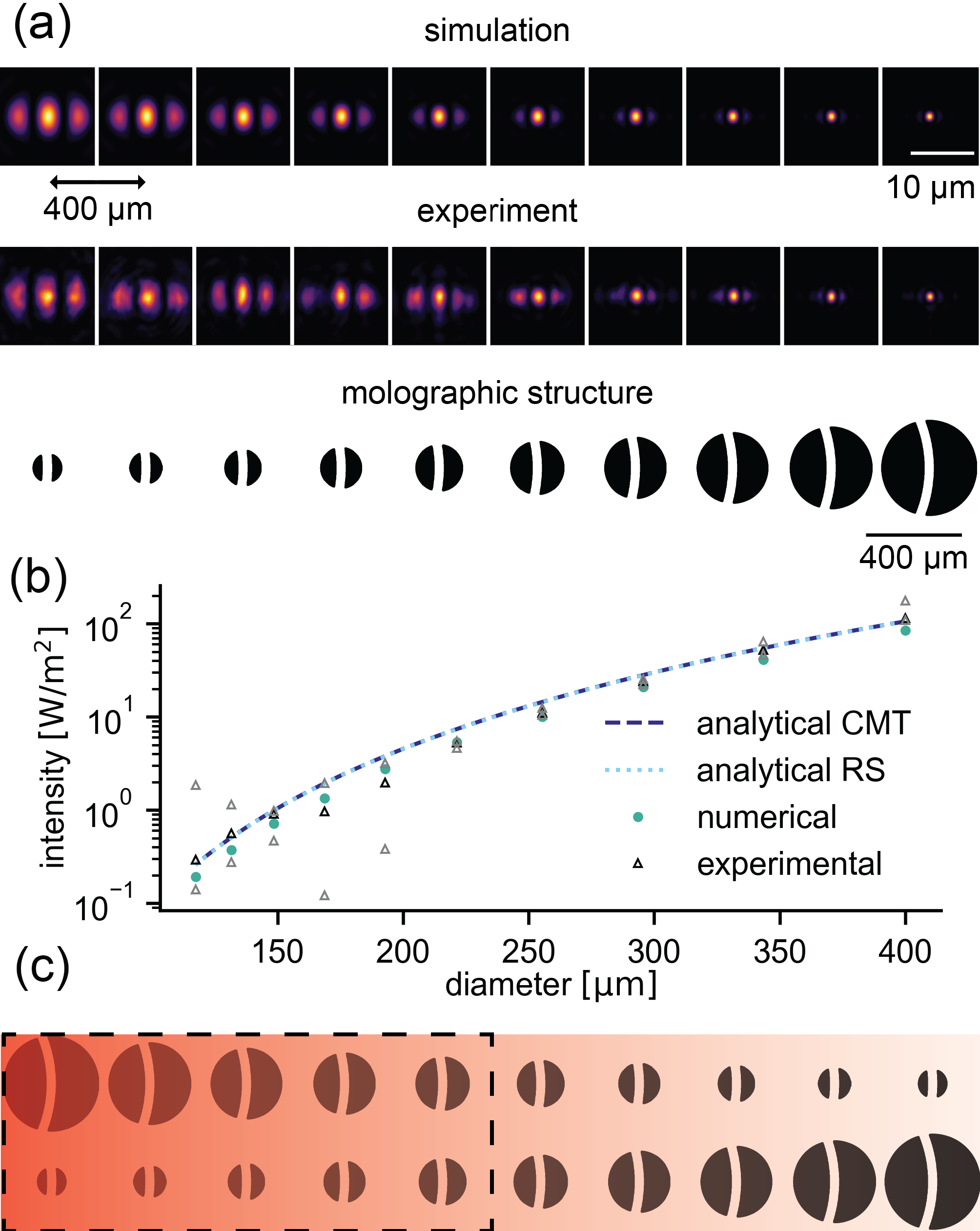}
\caption{\label{fig4}Comparison between analytical models [CMT and RS], numerical simulations and experiments for SAv mologram in air. (a) First row: Simulation of the intensity distribution in the focal plane for molograms with constant focal length [900 $\upmu$m], constant sinusoidal surface mass density modulation [540 pg/mm$^2$, peak-to-peak] and varying diameters with air as cover medium. Second row: Corresponding experimental measurement of the intensity distribution in the focal plane. Third row: Schematic of the underlying molograms. (b) Absolute average intensity values over the Airy disk for different molograms derived from measurements [median is in black other measurements in grey], simulations and the two analytical models [Eq. \eqref{I_sg_CMF} and \ref{I_sig_RS}]. The diameters of the molograms in the analytical formula were adjusted in order to account for the missing binding sites in the central Bragg recess area. The small differences between measured and calculated intensity for some molograms can be explained by alterations of the wave front of the guided mode due to preceding molograms (c) Chip geometry with the molograms positions taken for the analysis.} 
\end{figure}

The expression for Rayleigh scattering [neglecting the optical interfaces] is stated by Fattinger \cite{Fattinger2014-dm}. It can be written in the following form [see SM Section  \ref{sec:Molographic_signal_intensity_from_coupled_mode_theory_and_Rayleigh_scattering}]

\begin{equation}
	\label{I_sig_RS}
	{I_{{\text{avg,RS}}}} = 1.268 \cdot {\pi ^2}{\text{N}}{{\text{A}}^2}n_{\text{c}}^2\frac{{{{\left( {{n_{\text{P}}^2} - {n_{\text{c}}^2}} \right)}^2}}}{{{{\left( {n_{\text{P}}^2 + 2n_{\text{c}}^2} \right)}^2}}}\frac{{{D^2}}}{{{\lambda ^4}}}\frac{{{\eta _{{\text{mod}}}}{{_{\left[ {\text{A}} \right]}}^2}{\Delta _\Gamma }^2}}{{{\rho _{\text{P}}}^2}}{I_0}
\end{equation}
If one inserts the definition of the refractive index increment for proteins in dilute solutions [$\frac{dn}{dc}=0.182$ ml/g [water]] \cite{Zhao2011-ju,Heller1965-me}

\begin{equation}
\label{eq:ref_index_increment}
{{dn} \over {dc}} = {3 \over 2}{1 \over {{\rho _{\rm{P}}}}}{n_{\rm{c}}}{{n_{\rm{P}}^2 - n_{\rm{c}}^2} \over {n_{\rm{P}}^2 + 2n_{\rm{c}}^2}}
\end{equation}
one can easily verify that they yield the same result. $\rho_{\rm{P}}$ is the dry mass density of the protein calculated according to \cite{Fischer2004-io} and $n_{\rm{P}}$ the refractive index of the dry protein sphere. For SAv, we computed the following values: $\rho_{\rm{P}}=1.412$ $\rm{g/cm}^3$ and $n_{\rm{P}}=1.598$, which yield an equivalent $\frac{dn}{dc}=0.36$ ml/g [air] that has to be used in the CMT model for comparison purposes.

To compare the analytical expressions [Eqs. \eqref{I_sg_CMF} and \eqref{I_sig_RS}] with experiments, molograms of different diameters and numerical apertures were fabricated on a chip. 10 molograms of decreasing diameters [400, 343, 296, 255, 221, 193, 168, 148, 131, 117 $\upmu$m] and numerical apertures of [0.33, 0.29, 0.25, 0.21, 0.19, 0.16, 0.14, 0.13, 0.11, 0.1] at a constant focal length of 900 $\upmu$m were designed in the first row of the phase mask. The same 10 molograms were arranged in the opposite order in the second row of the phase mask. The diameters were chosen such that the area differs by a factor 1.4 from one mologram to the next [compensated for the Bragg recess area]. To use the analytical expression for molograms with the Bragg recess area the area has to be corrected by a factor $\frac{{2{A_ + }}}{{{A_{{\text{mologram}}}}}}$ where $A_+$ is the area of the ridges and $A_{\rm{mologram}}$ is the area of the molographic footprint [ridges + grooves + Bragg recess area]. The experimental design of two rows was chosen because each mologram alters the mode shape slightly and therefore the foci of the last molograms in a row are increasingly distorted in y-direction. Thus, the first five molograms of either row were used on three different measurement fields on the same chip [see Fig. \ref{fig4}(c) and Ref. \cite{,Gatterdam2017-oo} for chip geometry]. The investigated molograms were the same SAv-molograms as described in the last section. 

The simulation was performed by placing exactly the amount of SAv molecules sinusoidally on the ridges of the mologram that exhibits the same diffraction efficiency as a peak-to-peak surface mass modulation of 540 pg/mm$^2$ [for the 400 $\upmu$m mologram these are 231 million individual dipole scatterers, see SM Section \ref{sec:description_of_the_simulation_framework} for the necessary conversions]. The proteins were placed directly on the waveguide [the field at z = 0 was used to calculate the dipole moment]. The scattered intensity was computed on a 150 x 150 grid around the focus whereas individual grid points were spaced 110 nm apart, which is equivalent to the pixel size of the camera used in the experiment. Fig.  \ref{fig4}a shows the intensity distribution in the focal spot of the 10 molograms with varying diameters in air obtained from simulations and experiments, the last row shows the underlying mologram. It can be seen that the intensity distributions of simulation and experiment are in perfect agreement over the entire range of diameters investigated. It has to be noted, that it is nontrivial to achieve diffraction-limited focusing for molograms up to a diameter of 400 $\upmu$m on high refractive index waveguides, since already small gradients in the thickness [and therefore also in the effective refractive index] can cause an accumulated phase shift between the guided mode and the synthetic hologram [designed for constant effective refractive index]. Therefore, after a certain propagation distance, which we call the dephasing length, light scattered at the first and the last line of the mologram can interfere destructively [see SM Section \ref{sec:Dephasing_as_a_limitation_on_mologram_size} and Fig. \ref{fig:Dephasing}].

Fig.  \ref{fig4}(b) compares the average intensity in the focal point analytically [RS and CMT], numerically and experimentally. The numerical intensity values were obtained by averaging the intensity on the screen over the Airy disk of the diffraction-limited lens. Experimentally, the mean intensity was calculated by subtracting the average background and then averaging on a circle of the size of the Airy disk centered at the maximum intensity. As can be readily seen, the CMT model and the RS model show nearly perfect agreement with the experimental results. However, there are a few effects which are not accounted for in these simple analytical models. These include free space attenuation [all the dipoles are assumed to be at the center of the mologram], the angle dependence of Rayleigh scattering \cite{,Sinclair1947-xa}, the symmetry breaking of the central Bragg recess area, the observation in a half space with a denser medium and reflections at the interfaces \cite{,Lukosz1981-js}. The numerical simulations incorporate them [see SM Section \ref{sec:description_of_the_simulation_framework}], which result in a slightly lower intensity than the analytical models [factor 1.33 in air]. The fact that the experiment agrees closer with the simple analytical models can be explained by uncertainties in the measurement. These arise for example from the determination of the surface mass modulation on the mologram. Due to the nature of the quantification procedure [quantitative fluorescence on the sub-micron scale using STED], we expect to have an uncertainty of roughly 10 \% in this measurement. Other possible sources of error, and most likely the prominent ones, are the estimation of the guided power $P_{\rm{wg}}$ at the mologram location [SM Section \ref{sub:Power_calculations_for_the_molograms_in_Fig}]. One can also see in Fig.  \ref{fig4} that the molograms closer to the in-coupling grating have higher intensities and their median values show less deviation to the curves expected from the analytical and numerical models. This can readily be explained by the alteration of the wave front [guided-guided mode coupling] at every preceding mologram and at waveguide imperfections.

In summary, we have shown that the RS and CMT models are equivalent and show excellent agreement with numerical simulations and experiments for molograms with different diameters and numerical apertures in air. The analytical models are therefore a valid tool to make predictions of the limit of detection and to determine the surface mass modulation from the measured intensity in the molographic focal spot. Furthermore, we demonstrated the manufacturing of diffraction-limited molograms with diameters up to 400 $\upmu$m on a high refractive index waveguide. Using much larger molograms [>1 mm diameter] is not reasonable, since expressed proteins are valuable and often limited in biological experiments \cite{,Structural_Genomics_Consortium2008-hk}.



\section{Background and noise analysis for molography with massless affinity modulation} 
\label{sec:background_and_noise_analysis}

 Besides the intensity that originates from coherently arranged molecules on the waveguide, various background sources scatter intensity into the area of the focal spot.  This can either obscure the coherent signal or limit its accuracy due to the stochastic variation of the background. We analyzed the limit of resolution for the important case of molograms with massless affinity modulations. For such molograms the signal of the empty mologram is hidden in the speckle background. Fattinger \cite{,Fattinger2014-dm} provided a first estimation of the limit of detection by comparing the power diffracted by the mologram to the background power incident on the Airy disk. The background power was estimated by distributing the waveguide radiation loss uniformly over the solid angle [4$\pi$]. While this serves as good first approximation, we now refine the approach. First, we distinguish between background and noise. Whereas we describe the background as the mean intensity in the focal plane, we consider its spatial and/or temporal fluctuations as noise. Although the noise determines the limit of detection, it is worthwhile to investigate the background because the amplitude of the noise is in a fixed ratio to the mean background intensity. This is due to the nature of the speckle pattern \cite{,Dainty1977-nh}, which will be explained in more detail below.

\begin{figure}
\includegraphics{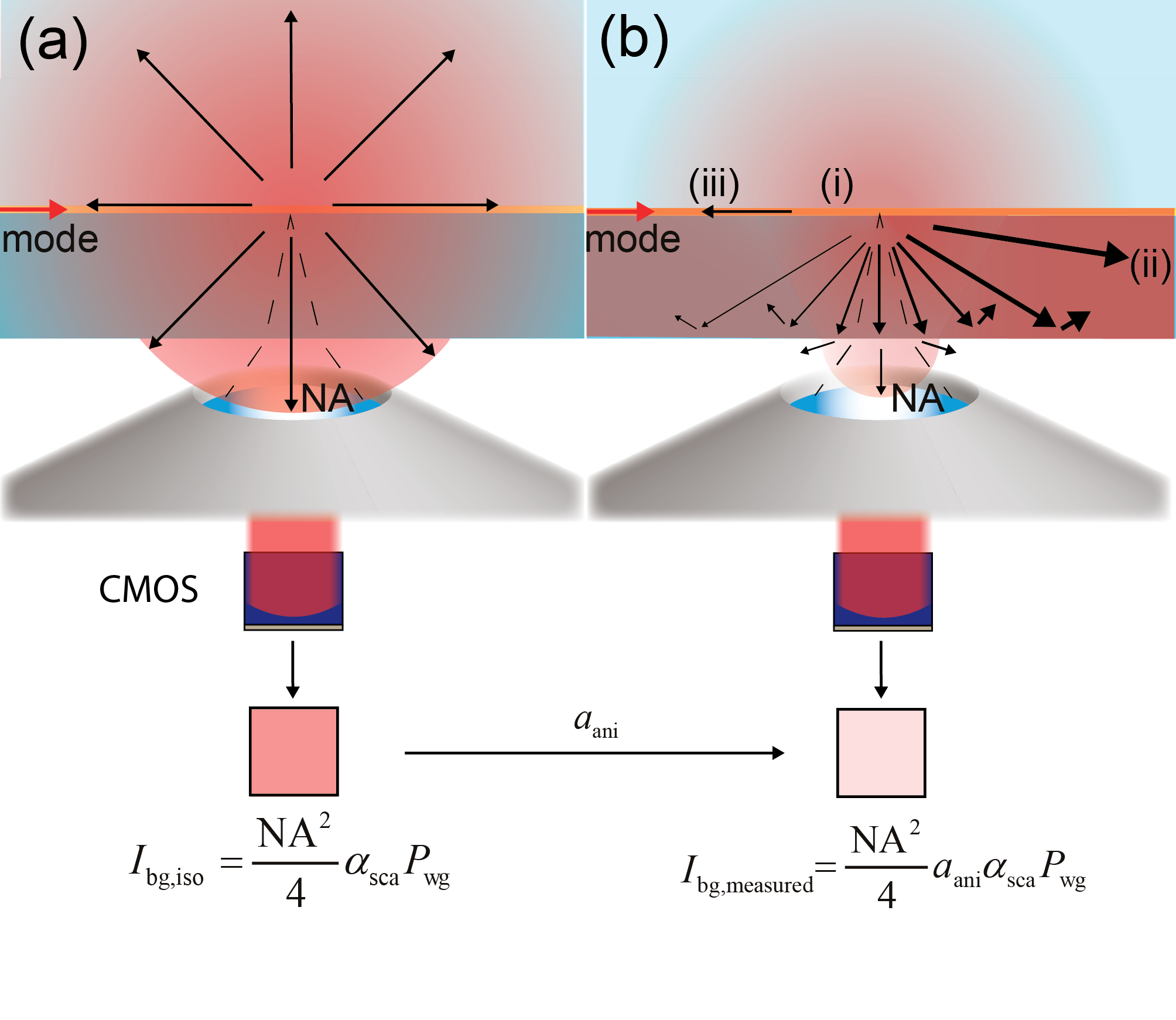}
\caption{\label{fig5} Illustration of the anisotropy parameter $a_{\rm{ani}}$. (a) The scattered power is distributed isotropically in all directions. Only angles that can be collected by the numerical aperture [dashed lines] of the objective contribute to the background. This results in the expression for $I_{\rm{bg,iso}}$. The intensity is then multiplied with $a_{\rm{ani}}$ to match the average measured intensity in the focal plane [see Fig. \ref{fig:ansiotropic_experimental}] (b) In reality, scattering is an anisotropic process. There are three effects that contribute to anisotropy. (i) The asymmetry of the waveguide leads to a stronger scattering into the substrate due to its higher optical density \cite{Novotny2011-pa}. (ii) Forward scattering is usually dominant over backward scattering \cite{,Miyanaga1980-od}. (iii) Scattering into guided modes of the waveguide is more efficient than into freely propagating modes \cite{Novotny2011-pa}.}

\end{figure}

The propagation loss [or attenuation] of a dielectric optical waveguide is an important quantity for its characterization. We evaluate the background with the help of the radiation loss as it has been performed in Ref. \cite{,Fattinger2014-dm} for a first estimation of the limit of detection. However, two issues arise when approximating the background from the propagation loss. First, the attenuation constant is a sum of absorption and scattering loss $\alpha = \alpha_{\rm{abs}} + \alpha_{\rm{sca}}$, where we define the propagation loss as ${P_{{\text{wg}}}}\left( x \right) = {P_{{\text{wg}}}}\left( 0 \right){e^{ - {\alpha}x}}$. The scattering loss provides additional background photons to the area of the focal spot, whereas the absorption loss does not contribute any additional light. Therefore, determining the background intensity with the propagation loss is only possible when the absorption is small compared to the scattering. The second issue when estimating the background from the attenuation arises from the anisotropy of the scattering. The out-coupled power is not distributed isotropically over the solid angle. In order to determine the intensity in the focal plane, an additional parameter is needed to account for the anisotropy. This anisotropy parameter $a_{\rm{ani}}$ is explained in more detail in Fig. \ref{fig5}. The average intensity of the background in the focal plane $I_{\rm{bg}}$ can be conveniently written in terms of the scattering loss and the anisotropy parameter [detailed derivation in the SM Section \ref{sub:Approximative_formula_for_the_background_intensity_in_the_focal_spot}].

\begin{equation}
	\label{I_FP}
	{I_{{\text{bg}}}} = \frac{{{\text{N}}{{\text{A}}^2}}}{4}{a_{{\text{ani}}}}{\alpha _{{\text{sca}}}}{P_{{\text{wg}}}}
\end{equation}
We call the product $a_{\rm{ani}}\alpha_{\rm{sca}}$ the \textit{scattering leakage}, since it refers to the light leaking into the focal plane of the mologram.

\begin{figure}
	\includegraphics{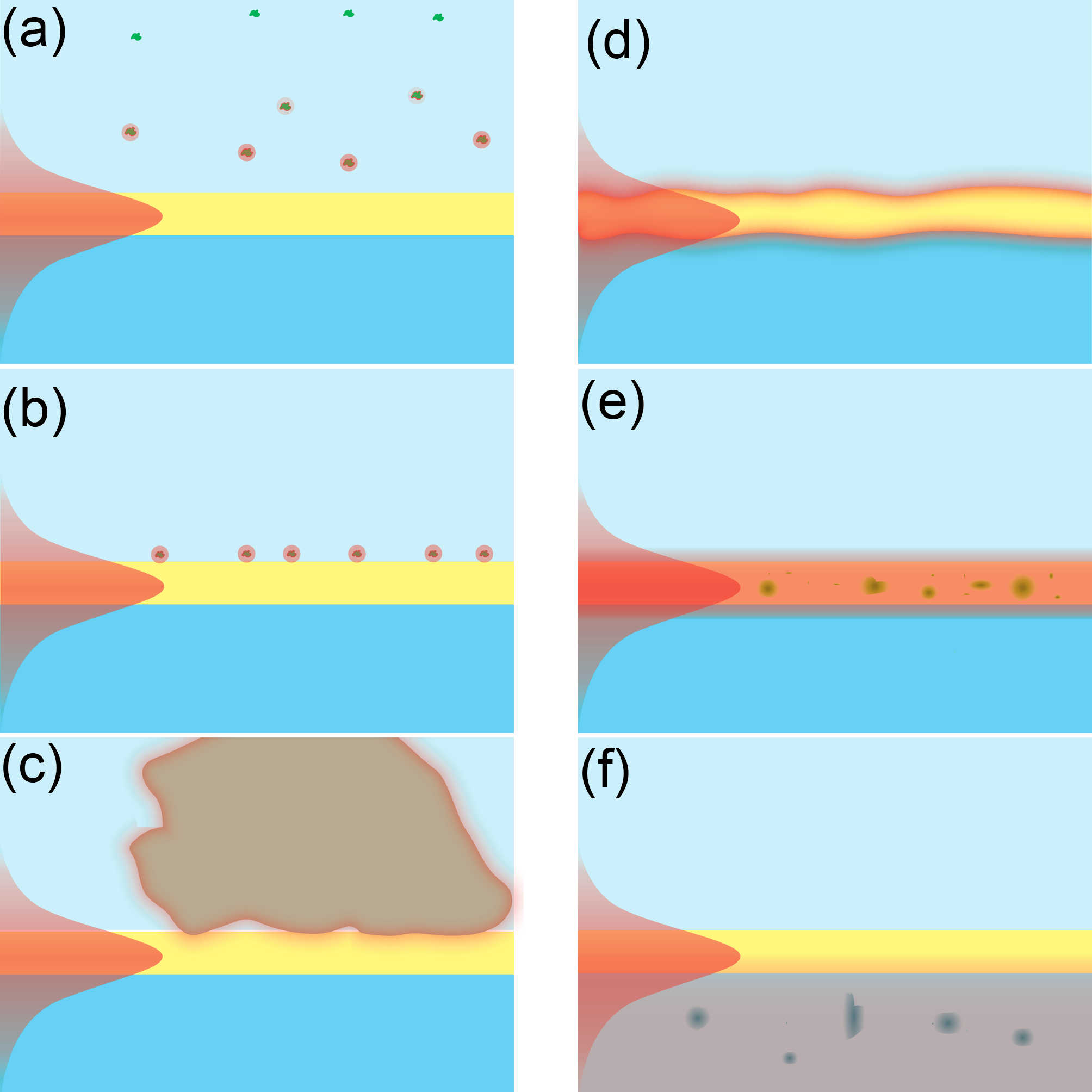}
	\caption{\label{fig6} Possible sources of background intensity due to scattering (a) molecules in the solution (b) non-specifically bound molecules on the waveguide surface (c) large particles such as dust or cell bodies on the waveguide surface (d) waveguide surface roughness (e) refractive index inhomogeneities inside the waveguide (f) refractive index inhomogeneities inside the substrate} 
\end{figure}

Next, we aim to show which scattering mechanisms exist, how they contribute to the background and what needs to be considered when designing a waveguide for molographic sensing. We assume these mechanism to be non-correlated, which allows us to add the individual contributions such that
\begin{equation}
	\label{sca_leak}
	{a_{{\rm{ani}}}}{\alpha _{{\rm{sca}}}} = \sum\limits_i {{a_{{\rm{ani,}}i}}{\alpha _{{\rm{sca,}}i}}}
\end{equation}
where the total scattering leakage $a_{\rm{ani}}\alpha_{\rm{sca}}$ corresponds to the experimentally measured value [see SM Section \ref{sub:Experimental_determination_of_scattering_leakage_and_anisotropy_of_scattering}]. Fig.  \ref{fig6} shows six possible scattering sources for background photons: (a)-(c) are scattering processes inherent to the biosensing experiment and (d)-(e) depend on the waveguide manufacturing. We will now investigate the importance of each mechanism qualitatively. If possible, we will treat the relevant sources quantitatively. Before analyzing each scattering process individually, the relation between background and noise shall be explained. Any scattering is caused by an underlying stochastic refractive index distribution within the angles of the numerical aperture of the optical system. This stochastic process is transformed by the coherent illumination to spatial intensity fluctuations in the focal plane - a speckle pattern. It is of utmost importance to distinguish between dynamic and static scattering processes. The speckle pattern of a dynamic scattering process exhibits a timescale much shorter than the required bandwidth of the sensor. It will be averaged to a homogeneous background. This background can be subtracted, which renders all dynamic scattering processes negligible. Conversely, static scattering backgrounds lead to speckle patterns that are relatively stable over the time course of the measurement and generally unknown \textit{a priori} to the measurement. This generates an uncertainty when we determine the mass density on the mologram because the relative contribution of the static background to the intensity of the molographic focus is unknown. As a side note, if the numerical aperture of the objective and the mologram match, background speckles and the mologram focus have the same size. Fortunately, the statistics of speckle patterns are well-known and speckles exhibit a negative exponential distribution of the intensity \cite{,Dainty1977-nh}. The 99.7 quantile [Definition of the LOD, generally stated as $\mu + 3\sigma$] of the exponential distribution is always in a fixed ratio to the mean and therefore knowing the mean background allows estimating the noise and the limit of detection.

Fig.  \ref{fig6}(a) illustrates the scattering at molecules in solution in the evanescent field above the waveguide. If the distance between two molecules changes by roughly half a wavelength, the interference for a given speckle in the focal plane can switch from completely constructive to completely destructive. When comparing this short distance to the diffusivity of proteins \cite{,Zhang2007-bb} it is apparent that the measurement time [approximately 1 s] is several orders of magnitude longer than the diffusion time over these length scales. Therefore, the scattering of any molecule in solution is a dynamic process and does not contribute to the noise in the background.

The scattering of randomly adsorbed proteins [Fig. \ref{fig6}(b)] on the waveguide surface has static and dynamic components. Most of these molecules adsorb reversibly to the low-energy non-fouling surface and therefore have affinities in the mM range. Interactions with such affinities exhibit short lifetimes [$\upmu$s-ms] \cite{,Liu2012-sx}. Therefore, the rate at which new proteins adsorb and desorb on the surface is much faster than the acquisition of a single data point. On the other hand, a minority [below 10 pg/mm$^2$] \cite{,Pasche2003-mr} binds quasi irreversibly to incoherent surface defects present on any monolithic surface ad-layer \cite{,Love2005-oq}. These are the static components of nonspecific binding. However, the contribution is extremely weak compared to the coherent signal. This can easily be understood if one recalls that the coherent signal scales with the number of adsorbed particles squared whereas the nonspecific background scales linearly with this number \cite{,Fattinger2014-dm}. To give an example, the 10 pg/mm$^2$ of irreversibly bound molecules upon serum exposure correspond to roughly 100 million molecules per mm$^2$. The same signal intensity can be achieved with a coherent arrangement of the square root of this number, which is about 10,000 molecules per mm$^2$. This corresponds to only 1 fg/mm$^2$ of coherent matter or four orders of magnitude less than the molecular mass from irreversible nonspecific binding. This consideration exemplifies again the insensitivity of focal molography to nonspecific binding \cite{,Gatterdam2017-oo}. The value is so low that in most applications other background sources will be limiting. 

The scattering originating from large particles [see Fig.  \ref{fig6}c] will be static or dynamic depending of the flow conditions. The analytical treatment of this mechanism is  difficult because such particles exceed the Rayleigh regime and the evanescent field of the waveguide. However, one can at least state that the large size of these scatterers causes strongly anisotropic forward scattering [Mie scattering regime \cite{,Hahn2009-sp}]. Besides the strong anisotropy, the influence of this scattering process can be reduced by controlling the number of adsorbed particles through careful handling of the chips and filtering of the samples prior to analysis. We will therefore not cover this scattering process in our theory since for most experiments it can be minimized to a negligible level [see Fig. \ref{fig:Relative_Imporance_Dust}].

The static scattering process at the surface roughness of the two waveguide sidewalls [Fig. \ref{fig6}(d)] has been studied extensively and is not negligible \cite{Marcuse1974-uo,Ladouceur1996-of,Payne1994-tw,Ames1983-bn,Miyanaga1979-ht,Ronald_Louis1992-qj}. In order to quantify the scattering we have adapted the analytical formulas derived by Lacey and Payne \cite{Lacey1990-ii,Payne1994-tw} for the more general case of an asymmetric waveguide [see SM Section \ref{sub:Sidewall_Roughness_Scattering}]. This model has the advantage of providing an analytical solution for the attenuation constant [Eq.  \eqref{final_damping_constant_scattering}]. However, more relevant for our analysis is the expression for the scattering leakage caused by the roughness of the waveguide.

\begin{eqnarray}
\label{scattering_leakage}
	{a_{{\rm{ani,r}}}}{\alpha _{{\rm{sca,r}}}} &=& {{{\pi}{n_{\rm{s}}}} \over {{\rm{NA}}}}{{\pi\left( {n_{\rm{f}}^2 - {N^2}} \right)} \over {{t_{{\rm{eff}}}}\left( {n_{\rm{f}}^2 - n_{\rm{c}}^2} \right)}}{{{\sigma ^2}} \over {{\lambda ^2}N{n_{\rm{f}}}}}\nonumber\\
	&&\times \left( {{{\left( {n_{\rm{f}}^2 - n_{\rm{c}}^2} \right)}^2} + {J^2}{{\left( {n_{\rm{f}}^2 - n_{\rm{s}}^2} \right)}^2}} \right)\\
	&& \times {L_c}\beta \int\limits_{{\frac{\pi}{2} -\frac{\rm{NA}}{n_{\rm{s}}}}}^{\frac{\pi}{2} + \frac{\rm{NA}}{n_{\rm{s}}}} {{1 \over {1 + {{\left( {\left( {\beta - k{n_{\rm{s}}}\cos \theta } \right){L_{\rm{c}}}} \right)}^2}}}} d\theta\nonumber
\end{eqnarray}

where $J$ accounts for the waveguide asymmetry.
\begin{eqnarray}
\label{asymmetry_parameter_of_the_waveguide}
J &=& \cos \left( {{t_{\text{f}}}\beta \frac{{{{\left( {n_{\text{f}}^2 - {N^2}} \right)}^{1/2}}}}{N}} \right) \\
&& + \frac{{{{\left( {{N^2} - n_{\text{c}}^2} \right)}^{1/2}}}}{{{{\left( {n_{\text{f}}^2 - {N^2}} \right)}^{1/2}}}}\sin \left( {{t_{\text{f}}}\beta \frac{{{{\left( {n_{\text{f}}^2 - {N^2}} \right)}^{1/2}}}}{N}} \right)\nonumber
\end{eqnarray}
$\sigma$ is the the RMS roughness and $L_{\rm{c}}$ the correlation length of the roughness. $\beta=\frac{2\pi N}{\lambda}$ corresponds to the momentum of the mode in propagation direction and $t_{\rm{f}}$ is the thickness of the waveguide. Eq. \eqref{scattering_leakage} can be used to estimate the background for different parameters. Yet, since it is difficult to determine a single correlation length from an AFM measurement the model should be applied carefully. It can be improved by taking into account the full information of the power spectral density instead of assuming an exponential decay of the autocorrelation. Yet, this model is less intuitive. In addition, the assumption of no correlation between the two sidewall roughnesses and the two-dimensionality of the model can result in some inaccuracy. Despite these simplifications, the described model is a valuable tool for the estimation of the background intensity from the roughness properties of the waveguide surface [see SM Section \ref{sub:AFM_characterization_of_surface_roughness}]. Furthermore, it is important to notice that the scattering leakage caused by roughness  $a_{\rm{ani,r}}\alpha_{\rm{sca,r}}$ is fairly constant with respect to NA. Once characterized, this enables a straightforward comparison between experiments of molographic system with different numerical apertures.

The volume scattering due to refractive index inhomogeneities in the waveguiding film [Fig. \ref{fig6}(e)] is a static process and amenable by similar theoretical investigation as the surface scattering \cite{Miyanaga1979-ht}. However, the applicability of these models are limited since the characterization of the detailed distribution of those inhomogeneities by an orthogonal method is not trivial. For our waveguides the volume scattering is much smaller than the scattering from surface roughness of the waveguide sidewalls [SM section \ref{sub:Relative_importance_of_surface_to_volume_scattering}] and can therefore be neglected. Yet, this holds only true for thin waveguides with a high refractive index. 

Finally, the scattering from substrate inhomogeneities can be neglected because it is extremely low for a well-chosen material [Fig. \ref{fig6}(f)].

Based on the qualitative analysis we conclude that for a general waveguide one has to consider the volume and surface scattering of the waveguide for the background analysis $a_{{\rm{ani}}}{\alpha _{{\rm{sca}}}} = {a_{{\rm{ani,r}}}}{\alpha _{{\rm{sca,r}}}} + {a_{{\rm{ani,v}}}}{\alpha _{{\rm{sca,v}}}}$. Which of these processes is prevailing is determined by the configuration of the waveguide. In general, the relative importance of volume scattering increases with waveguide thickness [see SM Section \ref{Influence_of_waveguide_parameters_on_signal_and_background}] On the other hand, the higher refractive index of the film, the larger is the index contrast and the light intensity at the two waveguide sidewalls, which increases the contribution of surface scattering. Therefore, surface scattering is likely to be dominant for thinner waveguides with a high refractive index contrast [which is the case for the Ta$_2$O$_5$ waveguide discussed in this publication, see SM Section \ref{Influence_of_waveguide_parameters_on_signal_and_background}], whereas volume scattering will be limiting for thick waveguides with a low refractive index [high fraction of power in the waveguide and hardly any index contrast at the waveguide sidewalls]. 

Yet, also the signal scales with the waveguide properties. Therefore, instead of minimizing the background one needs to maximize the signal to background ratio. We define a figure of merit for a waveguide in order to easily identify the relevant parameters for this optimization. More generally, a figure of merit of a molographic biosensor can be formulated. The figure of merit for focal molography is the ratio of signal to background intensity, which stands for the ratio of mass sensitivity to dark-field illumination quality.

\begin{equation}
\label{eq:figure_of_merit_focal_molography}
{\text{FO}}{{\text{M}}_{{\text{FM}}}} = \frac{{{D^2}}}{{{\lambda ^4}}}\frac{{{n_c}\left( {n_{\text{f}}^2 - {N^2}} \right)}}{{N{t_{{\text{eff}}}}\left( {n_{\text{f}}^2 - n_{\text{c}}^2} \right){a_{{\text{ani}}}}{\alpha _{{\text{sca}}}}}}
\end{equation}
This expression was obtained by dividing equation \eqref{I_sig_RS} by \eqref{I_FP}. We omitted the protein related parameters and the analyte efficiency since for most applications these are fixed. By further excluding the diameter of the mologram [geometrical design parameter], one obtains the figure of merit of the waveguide
\begin{equation}
	\label{figure_of_merit}
	{\text{FO}}{{\text{M}}_{{\text{WG}}}} = \frac{{{n_{\text{c}}}\left( {n_{\text{f}}^2 - {N^2}} \right)}}{{{\lambda ^4}N{t_{{\text{eff}}}}\left( {n_{\text{f}}^2 - n_{\text{c}}^2} \right){a_{{\text{ani}}}}{\alpha _{{\text{sca}}}}}}
\end{equation}
The dependency on wavelength to fourth power can be misleading. The choice of wavelength heavily affects the scattering leakage $a_{\rm{ani}}\alpha_{\rm{sca}}$. Other parameters such as $N$ and $t_{\rm{eff}}$ also depend on the wavelength. Therefore, one should be careful when making predictions from Eq. \eqref{figure_of_merit}. Experimentally, ${a_{{\rm{ani}}}}{\alpha _{{\rm{sca}}}}$ can be determined by measuring the intensity in the focal plane and the power in the waveguide and applying Eq. \eqref{I_FP}. Alternatively, $a_{\rm{ani}}\alpha_{\rm{sca}}$ can be estimated from the measured roughness properties with the help of Eq. \eqref{scattering_leakage}, if volume scattering is negligible compared to surface roughness scattering [see SM Section \ref{sub:AFM_characterization_of_surface_roughness}]. The relative importance of volume to surface scattering can be investigated by measuring the ratio of the scattered intensity with two different cover media, because volume scattering will be hardly affected by the change in cover medium. We performed these characterization for our waveguide and found a scattering leakage of $a_{\rm{ani,r}}\alpha_{\rm{sca,r}}=3.12$ /m in air and surface roughness scattering to be dominant over volume scattering [see SM Section \ref{sub:AFM_characterization_of_surface_roughness}]. From this we computed a figure of merit of our waveguide of $\rm{FOM}_{\rm{WG}}=1.27\cdot10^{30}$ /m$^4$. The figure of merit for molography for a mologram of diameter 400 $\upmu$m on this waveguide is $\rm{FOM}_{\rm{FM}}=2\cdot10^{23}$ /m$^2$.

In summary, the current high refractive index Ta$_2$O$_5$ waveguide is already a good choice for molography. Mainly thanks to the high field on the surface, which leads to a strong signal and also to its negligible volume scattering. This compensates for the fairly large surface scattering due to the substantial index contrast. Still, since surface scattering is dominant, we expect that the waveguide could be further optimized. The options to reduce the surface scattering include to diminish the RMS of the surface roughness, which was determined with atomic force microscopy to be 0.6 nm for our waveguide [see SM Section  \ref{sub:AFM_characterization_of_surface_roughness}], or to adapt the waveguide parameters using Eqs. \eqref{scattering_leakage} and \eqref{figure_of_merit}. However, this has to be performed with care, since Eq. \eqref{scattering_leakage} only considers surface roughness scattering. For other waveguides, such as thick low refractive index waveguides, volume scattering is most likely the dominant scattering source and the decrease in sensitivity for low refractive index waveguides is substantial. Nevertheless, once the scattering leakage is assessed experimentally, Eq. \eqref{figure_of_merit} allows a straightforward comparison of different waveguides.


\section{Limits of detection and resolution} 
\label{sec:limit_of_resolution_and_detection_of_focal_molography}

After having described the possible sources of background light and formulated the figure of merits we will use this knowledge to analyze the limits of detection and resolution of focal molography. The proper limit of detection for a specific assay is elaborate \cite{Armbruster2008-ol,Holstein2015-ej}. Yet, it is not practically feasible to compare different sensing platforms at the assay level since this would require a standard assays to be performed with each of them. To establish the detection limit in endpoint measurements for molography, we define it as a confidence level for false positives, namely the 99.7 percentile [for the normal distribution equal to $\upmu$ + 3$\rm{\sigma}$] or if not possible due to lack of experimental data points the 99 or 99.5 percentile. The resolution is a common benchmark parameter for sensors in general and for real-time label-free biomolecular interaction analysis in particular \cite{Homola2008-px}. It is defined as the temporal rms [root mean square] noise of the baseline after drift correction for the duration of a typical biosensing experiment. Next, we need to find a suitable unit to compare these two quantities amongst molography and other biosensors. This is not straightforward, since most biosensors measure a change in ad-layer density, yet, focal molography measures a change in ad-layer density modulation. In order to enable comparison,  we propose the molographic surface mass density $\Gamma$ to be this quantity for focal molography. It is defined as the entire mass in the mass modulation uniformly spread over the mologram [see SM Section \ref{sec:definitions}]. This quantity is calculated from the molographic signal intensity ${{I_{{\text{sig}}}}}$ normalized by a reference intensity. A suitable intensity reference for massless affinity modulations is the mean of the speckle background ${{I_{{\text{bg}}}}}$, since it is affected in the same manner as the molographic focus by the majority of physical processes that cause noise or drift [see SM Section \ref{sub:Intensity_reference_molography} for a discussion on intensity references]. Not in all bioanalytical questions it will be possible or simply too expensive to obtain completely massless affinity modulations [i.e. a small ligand interacting with a large immobilized protein]. In this case, a reference hologram will be required in order to calibrate the molographic signal in samples with varying bulk refractive index. This is not to be confused with the inherent self-referencing character of diffractometric sensors which makes them more robust than referenced refractometric sensors. Contrary to these, diffractometric sensors only measure the refractive index difference between ridges and grooves. Therefore, they do not need to compensate for refractive index drifts in the entire volume of the evanescent field. Finally, we need to specify what the molographic signal ${{I_{{\text{sig}}}}}$  refers to exactly. So far, we discussed the average intensity in the Airy disk, which is related to the maximum by $I_{\rm{max}}=4.378\cdot I_{\rm{avg}}$, as the potential molographic signal. Equation \eqref{eq:surface_mass_density} and the remainder of this paper use a different algorithm for the computation of the molographic signal which amounts to $I_{\rm{sig}}=2.012\cdot I_{\rm{avg}}$. ${{I_{{\text{sig}}}}}$ is the intensity on a single pixel of the focal plane image after convolution with a normalized Bessel kernel of the size of the expected Airy disk [see SM Section \ref{sub:Derivation_of_the_limit_of_detection_formula}]. The molographic surface mass density $\Gamma$ can directly be calculated from the ratio of the molographic signal and the mean intensity of the background as:

\begin{equation}
	\label{eq:surface_mass_density}
	\Gamma = {\Gamma _0}\sqrt {\frac{{{I_{{\text{sig}}}}}}{{{I_{{\text{bg}}}}}}}
\end{equation}
where $\Gamma_0$ [see SM Section \ref{sub:Derivation_of_the_limit_of_detection_formula} for derivation]
\begin{eqnarray}
	\label{gamma0}
	{\Gamma _0} &=& 0.1056 \cdot \frac{{{A_ + }}}{{{A_{{\text{mologram}}}}}{}}\frac{1}{{\eta _{{\text{mod}}}}_{\left[ {\text{A}} \right]}}\frac{1}{{\sqrt {{\text{FOM}}_{{\text{FM}}}}}\frac{dn}{dc}}
\end{eqnarray}
can be seen as an equivalent molographic mass density [calculated from the intensity in the focal plane by using the scattering strength of biological matter]. It is due to the stochastic variation of the significantly larger equivalent incoherent mass density of all the incoherent scatterers. The equivalent molographic mass density would contribute the same amount of intensity as the scattering leakage to the focal plane of the mologram [average background intensity in focal plane]. For the rest of this paper we assume a sinusoidal mass modulation [${{\eta _{{\text{mod}}}}_{\left[ {\text{A}} \right]}}=0.5$]. 

The limit of detection in terms of molographic surface mass density for a figure of merit of a given molographic system can be obtained by replacing ${{I_{{\text{sig}}}}}$ with $I_{\text{sig,LOD}}$ and $\Gamma$ with ${\Gamma _{{\text{LOD}}}}$ in Eq. \eqref{eq:surface_mass_density}. The minimal detectable normalized intensity increase  $\frac{\Delta I_{\text{sig}}} {{{I_{{\text{bg}}}}}}$ is determined by the readout scheme. We will now distinguish between two readout schemes: Endpoint detection and real-time measurements. We will derive the limit of detection and the limit of resolution for the two schemes,respectively. We also provide experimental evidence for our statements as well as the theoretical projection of the optimization potential.

\subsection{The limit of detection of endpoint measurements is determined by the statistics of the speckle background} 
\label{sub:use_case_1_detection_limit_in_a_speckle_background_without_any_pre_knowledge_of_the_speckle_intensity_distribution}

\begin{figure*}[t]
	\includegraphics{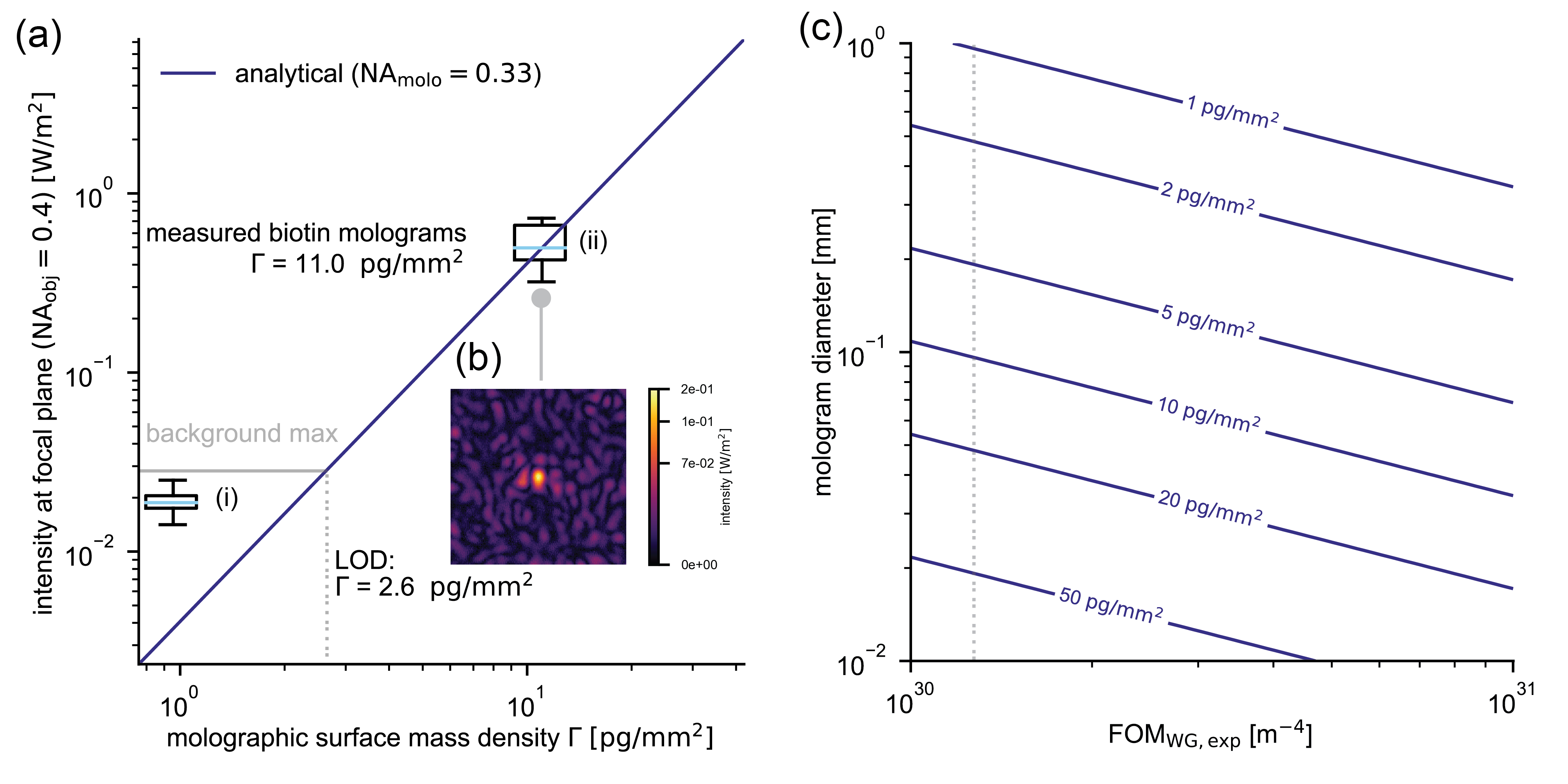}
	\caption{\label{fig7}Detection limit of focal molography without pre-knowledge of the position of the focal spot for the standard configuration in air [diameter mologram 400 $\upmu$m, focal length 900 $\upmu$m, numerical aperture of the mologram 0.33 and numerical aperture of the objective 0.4]. (a) Detection limit for the standard configuration mologram. 180 background images [280 x 210 $\upmu$m with 110 nm pixel size] focused 100 $\upmu$m below the surface of the waveguide of three clean chips with a scattering leakage of 3.12 /m [Figs. \ref{fig:Damping_Constant} and \ref{fig:ansiotropic_experimental}] were acquired, filtered with the shape of the Airy disk [NA = 0.33] and the maximum pixel of the convoluted image extracted and summarized in the box plot (i). All measured intensity values were normalized to a standard power of 0.02 W/m in the waveguide. The solid grey line corresponds to the 99 \% percentile of the maximum pixels observed in the Airy disk convoluted background images defining the smallest intensity required for the coherent signal to be discriminated against surface roughness speckles. The solid blue line represents the scattered intensity as a function of the molographic mass density [sinusoidal mass distribution and calculated from Eq. \eqref{I_sig_RS} with the mass modulation replaced by the molographic mass density]. The intersection point is indicated by the dashed grey line, which denotes the coherent mass that corresponds the 99 \% percentile of the measured maximum background intensity for a field of view as specified above. The box plot (ii) corresponds to 12 measured biotin molograms in air. (b) Typical focal plane image of easily detectable biotin molographic foci with intensities roughly 10 times above the detection limit. (c) Detection limit in terms of molographic surface mass density for an ideal sinusoidal mass modulation as a function of the figure of merit of the waveguide [$\mathrm{FOM}_{\mathrm{WG}}$] and the diameter of the mologram. The detection limit decreases inversely with the diameter and decreases inversely to the square root of the $\mathrm{FOM}_{\mathrm{WG}}$. The dashed grey line indicates the $\mathrm{FOM}_{\mathrm{WG}}$ of the investigated waveguide.
	}
\end{figure*}

 In an endpoint measurement the operator has usually no $\textit{a priori}$ knowledge of the intensity distribution of the speckles and the location of the focal spot [except the focal distance to the surface]. Therefore, the Airy disk needs a certain brightness compared to the background intensity to be detectable in a sufficiently large field of view. The detection limit is determined by the variation over many images of the ratio of the maximum pixel value to the image mean. In order to experimentally determine this value 180 images of size 280 x 210 $\upmu$m$^2$ with 110 nm pixel size on three different ZeptoMark [Zeptosens AG, Switzerland] chips were acquired and convoluted as described in SM section \ref{sec:Resolution_of_focal_molography_and_comparison_to_SPR_-_real-time_measurements}. The 99.5 \% quantile of this ratio equals to 13.8 [see Fig. \ref{fig:Speckle_Distributions}]. Hence, if the maximum pixel after convolution is 13.8 times brighter than the mean of the speckle background it is likely to stem from a coherent binding signal. By inserting this into Eq. \eqref{eq:surface_mass_density}, one obtains a detection limit in terms of molographic surface mass density of 2.6 pg/mm$^2$ for a 400 $\upmu$m mologram with 0.4 NA on our waveguide, which has a figure of merit of $1.27\cdot10^{30}$ /m$^4$. [Fig. \ref{fig7}(a)]. Therefore, only 336 fg of matter yield a signal that is clearly assignable to a coherent assembly of molecules. Fig. \ref{fig7}(c) displays a contour plot of the two parameters that affect the detection limit of a molographic system - the diameter of the mologram and the figure of merit of the waveguide. An increase of either of these two parameters decreases the detection limit. It should be mentioned that the endpoint measurements with no $\textit{a priori}$ knowledge of the speckle intensities represents an upper bound of the achievable detection limit of diffraction-limited molography. Any readout scheme of greater sophistication will achieve lower detection limits.

\subsubsection*{Endpoint detection of vitamin B7 [biotin]} 
\label{sub:detection_of_vitamin_b7_biotin_molograms_in_a_speckle_background}


In this section we demonstrate experimentally that molography can visualize a label-free mass modulation caused by a low molecular weight compound in an endpoint measurement. Molograms with the sole difference between grooves and ridges being the tiny molecule vitamin B7 [molecular weight 227 g/mol] were fabricated [NH-biotin|NH$_2$]. A chip was illuminated with a dose of 2000 mJ/cm$^2$, incubated with 1 mM sNHS-biotin in HBS-T buffer [10 mM HEPES, 150 mM NaCl, 0.05 \% Tween20] at pH 8.0 for 15 min and flood exposed as described in \cite{Gatterdam2017-oo}. The foci of these biotin molograms were easily detectable [Fig. \ref{fig7}(b)]. To prove that indeed biotin molograms and not just random speckles were measured, a real-time video of the focal point was recorded while the grooves were backfilled with biotin as described in the description of Movie 1. One can clearly see the intensity gradually decreasing until the focal spot becomes indistinguishable from the speckle background. The molographic mass density [sinusoidal modulation] calculated from the median intensity via Eq. \eqref{eq:surface_mass_density} of 12 molograms amounted to 11 pg/mm$^2$ $\pm$ 1.6 pg/mm$^2$ [$\mathrm{FOM}_{\mathrm{WG}}=0.58\cdot10^{30}$ /m$^4$, calculated from the attenuation constant assuming the same $a_{\rm{{ani}}}$]. The uncertainty is caused by an intrinsic property of speckle statistics. Every speckle has a non-negative intensity. However, the sign of the electric field of a background speckle can be positive or negative and is unknown, since the phase information cannot be measured. If the field of the speckle is negative with respect to the field of the molographic signal some additional coherent matter is required to cancel the contribution from the roughness. Vice versa, less biological mass is required in the case when the molographic focus happens to be on a positive speckle. This physical property poses an intrinsic constraint on the accuracy of the molographic measurement and therefore the LOQ [limit of quantification].

It has to be stressed that the detection of this low molecular weight compound was possible without any equilibration, referencing or stabilization of the sensor. The amount of bound mass was determined in a reproducible fashion with an accuracy below 2 pg/mm$^2$ even if the chip is removed and reinserted. This robustness is a fundamental difference to common refractometric sensors, where similar detection limits can only be achieved with samples that are mounted and stabilized within the device and cannot be removed and reinserted.

\subsection{The resolution of real-time measurements is determined by the temporal noise of the speckle background} 
\label{sub:use_case_3_in_the_speckle_background_with_consecutive_measurements_real_time_detection_}

\begin{figure}
\includegraphics{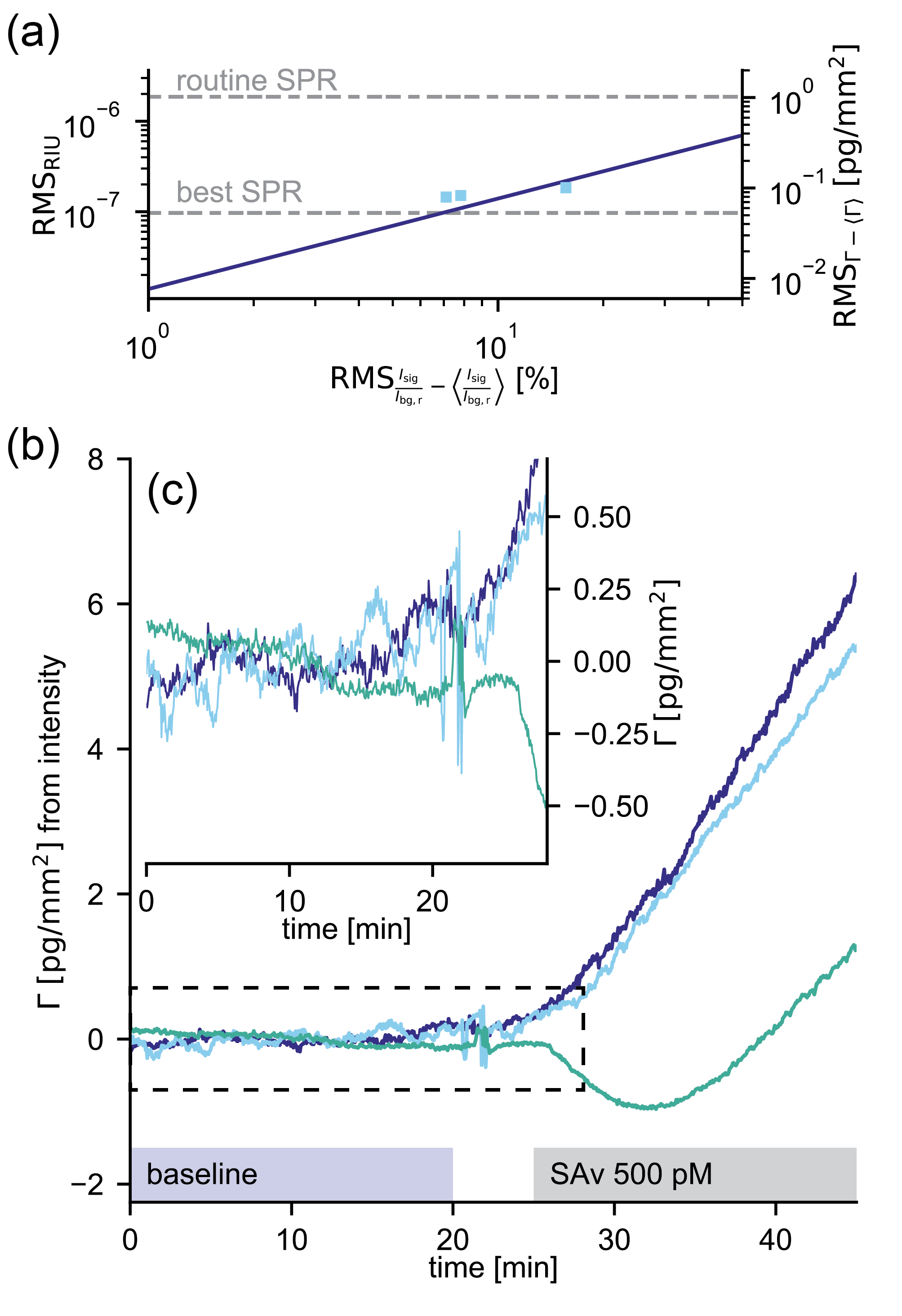}
\caption{\label{fig8}Comparison of the resolution of focal molography to state of the art resolution of SPR. (a) Best and routine resolution of SPR instruments is indicated.  Blue squares are the resolution measured for focal molography.  The dark blue line is the sensitivity of molography for a $\mathrm{FOM}_{\mathrm{FM}}$ of $10^{23}$ /m$^2$. (b) label-free detection of 500 pM [26 ng/ml] of SAv and baseline noise levels. (c) Zoom in of the first part of the binding curve for better visualization of the baseline noise.} 
\end{figure}

In real-time detection the intensity in the focal plane is continuously monitored. Contrary to the above described end-point measurement, the location of the focal spot in the speckle background is known exactly. This could be realized by reference focal spots or by localizing the spot before backfilling as shown below. Then image processing can be applied to monitor the intensity at the location of the focal spot resulting in a binding trace as a function of time.  Such a binding curve is the output of all real-time biosensors and the detection limit is commonly stated as the temporal RMS noise [resolution] over a defined time span, typically a few minutes, of the signal before the analyte is injected. It should be noticed that this is a different definition than the 99.5 \% quantile described above. The goal of this section is to demonstrate the resolution of diffraction-limited focal molography theoretically and experimentally for our measurement system and compare it to SPR, the gold standard of refractometric sensing.

We start our discussion by appreciating the instrumental precision at which refractometric sensors are operated in order to achieve refractive index resolutions of 10$^{-6}$-10$^{-7}$ and mass resolutions of 30 fg/mm$^2$ - 1 pg/mm$^2$ \cite{,Piliarik2009-xb,Homola2008-px}. If one recalls that one mono-layer of water molecules already gives rise to a signal of 300 pg/mm$^2$ one can appreciate of this technological achievement. Only careful optimization of sensor design, referencing strategies and signal processing over the past three decades made this possible \cite{Schasfoort2017-hs}. The performance of real-time measurement devices is commonly characterized by two metrics, namely the baseline drift  and the baseline noise. The former is expressed in pg/[mm$^2$min] or RU/min [RU = response units] whereas the latter is expressed as an RMS value in RU or pg/mm$^2$. Nowadays, commercial SPR instruments achieve a baseline noise of 15-30 fg/mm$^2$ [measured as RMS value after drift correction] and a baseline drift of around 300 fg/[mm$^2$/min] \cite{noauthor_undated-ex}. Over the course of a measurement SPR sensors are usually limited by temperature drifts between reference and sensing channel \cite{Kolomenskii1997-go}. As it will be shown below, such drifts are virtually not present in diffractometric sensors. Therefore we will compare molography to idealized SPR instruments which are limited by the baseline noise.

The three most common readout modes of SPR are angle interrogation, wavelength interrogation and intensity interrogation \cite{Homola2006-uz}. Independent of the interrogation mode, the readout of the SPR signal is a measurement of a relative intensity $I_{\rm{reflected}}/I_{\rm{in}}$. The noise in the intensity results in a noise of the detected surface mass density, which determines the resolution of the technique \cite{Homola2008-px}. The resolution can be calculated from the intensity noise and the sensitivity.

\begin{equation}
\label{IntensityNoiseConversion}
{\text{RMS}}_{\Gamma  - {\left<\Gamma\right>}} = {{\text{RMS}_{\rm{R} - \left<\rm{R}\right>}} \over {S_{\Gamma}}}
\end{equation}
The sensitivity of SPR is stated in SM Section \ref{sub:Derivation_of_the_sensitivity_of_SPR}. As described in Eq. \eqref{eq:surface_mass_density} molography also measures a relative intensity [$I_{\rm{sig}}/I_{\rm{bg}}$] such that Eq. \eqref{IntensityNoiseConversion} is valid. For molography, the sensitivity can be described by [see SM Section \ref{sub:Derivation_of_the_sensitivity_of_focal_molography} for derivation] 

\begin{eqnarray}
\label{eq:sensitivity_of_molography}
{S_{{\Gamma _{{\text{FM}}}}}}  = \frac{2}{{{\Gamma _0}}}\propto \sqrt{\rm{FOM_{\rm{FM\textbf{}}}}}
\end{eqnarray}
Due to the quadratic nature of the sensor transfer function Eq. \eqref{eq:sensitivity_of_molography} is only valid if the signal intensity is close to the reference intensity [in this case the background intensity]. It must be stated here that the values for the sensitivity of molography and SPR should not be compared since they depend on the chosen reference intensity [$I_{\rm{bg}}$ and $I_{\rm{in}}$]. Instead the resolution [${\text{RMS}}_{\Gamma  - {\left<\Gamma\right>}}$] can be used for comparison. With the MoloReader an intensity baseline noise of about 10 \% has been measured resulting in a resolution of 90 fg/mm$^2$ [0.9 RU] for the waveguides used in this experiment [FOM = $0.63\cdot10^{30}$ /m$^4$]. This is close to the above mentioned 30 fg/mm$^2$ [0.3 RU] resolution of the best SPR sensors \cite{Piliarik2009-xb}. If the waveguide background is the intensity reference, Eq. \eqref{eq:sensitivity_of_molography} can be used to improve the resolution of molography, which can be achieved by a higher $\rm{FOM_{\rm{WG}}}$ or by a larger diameter of the mologram. Further, any reduction of the intensity noise will also significantly improve it.

In order to verify these findings real-time measurements were performed with the MoloReader [The setup is illustrated in Fig. \ref{fig:Seatup_Realtime}.]. First, it has to be stressed that all reported experimental results are without any kind of temperature stabilization. The sole effect of temperature in a molographic measurement is a slow drift of the location of the molographic spot within the focal plane, but its intensity is hardly affected by the temperature drift [Movie 2]. This movement can easily be compensated for by simple image registration algorithms, which was implemented in the readout algorithm [Fig. \ref{fig:Data_Processing}] \cite{Thevenaz1998-lb}. As mentioned above, for a continuous measurement, the Airy disk must to be in the field of view and the optical system needs to be focused on the focal plane [see SM Section \ref{sub:localization of molographic foci} for the protocol to accomplish this]. Buffer baselines were acquired at a flow rate of 20 $\upmu$l/min for 20 mins with a syringe pump [NE-511L, PumpSystems Inc.] and 1 ml syringes [Henke Sass Wolf GmbH]. The buffer solutions were degassed prior to use, to avoid noise caused by micro bubble formation. The exposure time of the camera was set to 500 ms and an image was acquired every 3 seconds. After 20 minutes, the syringe was exchanged by another one containing 500 pM SAv in PBS-T [spikes in the binding trace around 22 min]. Injection was continued at a flow rate of 20 $\upmu$l/min. Finally, image processing was performed in order to obtain the baselines as described in SM Section \ref{sub:Processing_of_real-time_binding_signals} [Fig. \ref{fig:Data_Processing}].

Three binding curves were acquired and are displayed in Fig.  \ref{fig8}(b). The signal change due to SAv binding was detectable almost instantaneously after injection. However, whereas in two measurements the binding trace rose immediately after injection, it decreased at first in the third measurement. This is an example of the molographic focus lying on a background speckle with a negative electric field with respect to the focus itself, as explained before. The baseline noise over 20 min amounted to RMS values of 0.074, 0.094, 0.077 RU and in terms of normalized intensity 7.1, 15.6, 7.8 \%. These experimental noise values nicely agree with the theoretical prediction from Eqs. \eqref{IntensityNoiseConversion} and \eqref{eq:sensitivity_of_molography}. As mentioned before, the mass density resolution is comparable to the best reported SPR results \cite{Piliarik2009-xb}. Yet, the molographic baselines were calculated without any baseline drift correction unlike the common practice in refractometric sensing demonstrating the robustness and sensitivity of focal molography.   


Another fundamental requirement in label-free interaction analysis is the ability to detect a distributed ensemble of molecules on a sufficiently large area. In other words, to detect low receptor [capture molecule] occupancies, a fact overlooked by most of todays nanosensing and single molecular detection concepts \cite{Squires2008-pt}. In sensitive assays the concentration of the analyte is usually several orders of magnitude lower [10 fM - 1 pM] \cite{Surinova2011-uf} than the dissociation constant of the capture probe [10 pM - 1 nM] \cite{Landry2015-xe}, which leads to a receptor occupancy of typically 0.1-1\% \cite{Squires2008-pt}. The molographic focus of a 400 $\upmu$m diameter mologram monitors the activity of roughly 1 billion recognition sites continuously and is therefore able to resolve low receptor occupancies as well as measuring a sufficient number of analyte molecules. For example, at the demonstrated resolution of 100 fg/mm$^2$ [roughly 1 million SAv molecules per mm$^2$] 100000 proteins are bound to one biotin mologram [receptor density of $3\cdot 10^{10}$ molecules per mm$^2$ [11 pg/mm$^2$]]. Furthermore, taking into account that four biotin molecules bind one SAv molecule the receptor occupancy in the experiments shown in Fig. \ref{fig8} can be estimated to amount to only 0.01 \%.


\section{Conclusions} 
 \label{sec:conclusion}
 
In molography, the coherent arrangement of binding sites in the mologram and the resulting insensitivity to non-coherent noise sources enables robust and highly sensitive detection of biomolecular interactions. A quantitative analysis of these interactions is amenable through the analytical models presented in this paper. These models compute the amount of biological matter bound to the mologram from the intensity of the molographic focus. Their accuracy is proven by the excellent agreement with the presented numerical simulations and the discussed experiments. High sensitivity and a low background are achieved by a waveguide providing field enhancement and a proper dark field illumination. However, radiation due to scattering at waveguide imperfections remains the dominant source of background light for massless affinity modulations. Therefore, figures of merits were introduced to investigate the parameter dependencies of the signal to background ratio. They allow straightforward comparison of different molographic arrangements and waveguides. Two readout schemes, endpoint detection and real-time measurements prove the intrinsic robustness and high sensitivity of focal molography. In an endpoint measurement, the low molecular weight compound vitamin B7 could be easily detected and the limit of detection in terms of surface mass was just a few pg/mm$^2$ by this simple readout scheme. The more elaborate real-time measurements exhibited a resolution below 100 fg/mm$^2$ over 20 min without any drift correction. This is comparable to the best commercially available refractometric sensors. With further optimization, it is therefore likely that the resolution of diffractometric sensors will surpass the one of refractometric devices. Yet, by only detecting the coherent signal, the coherent detection scheme has unmet advantages over any established label-free biosensor. Its unique combination of robustness and high sensitivity will enable numerous new applications to analyze the interactions of biomolecules in their natural habitat - the crowded environments of body fluids, tissues, cells and membranes.

\begin{acknowledgments}
We would like to thank Lukas Novotny [ETH] for valuable input of the theoretical treatment of dipoles near interfaces; Arens Winfried [IMT Masken und Teilungen AG, Greifensee] for the fabrication of the SiO$_2$ covered chips; Louis Palavi [ETH] for programming the alpha version of the reader control software in course of his internship; René Rietmann [Roche] for machining the fluidic parts and Stephen Wheeler [ETH] for construction and machining of the reader stage system; and finally, Roland Dreyfus [ETH] for numerous discussions on coupled mode theory.
\end{acknowledgments}

\appendix

\section{Author contributions}

A.F., Y.B., J.V. and C.Fattinger planned the experiments, which were performed and evaluated by A.F. A.F and C.Fattinger designed the instrumentation. Y.B. and A.F derived the analytical expressions. S.B. and A.F implemented the simulation framework which was adapted for GPU computing by C.Forró. M.L. conducted the AFM measurements and V.G. provided the foundations for many experiments. A.F. and Y.B. wrote the manuscript with C.Fattinger providing input for the content and the structure of the work.

\section{Movie descriptions}

\paragraph*{Movie 1: Real-time backfilling of biotin molograms} 
\label{par:video_1}

This movie shows the real-time backfilling of a biotin mologram [NH-biotin|NH$_2$] with 1 mM sNHS-biotin at pH 8.0 in HBS-T buffer. The molographic spot fades away upon biotin binding because also the grooves are functionalized with biotin, essentially canceling the mass modulation [NH-biotin|NH-biotin]. This proves that our investigated molograms in Fig.  \ref{fig7} were indeed biotin molograms.

\paragraph*{Movie 2: Temperature effect on the molographic spot} 
\label{par:movie_2_temperature_effect_on_the_molographic_spot}

This video shows the influence of temperature on the speckles in the focal plane image. The speckles as well as the molographic spot shift as a function of temperature but their intensity essentially remains constant. The chip was observed in PBS-T buffer without any flow and the entire chip assembly [Fig. \ref{fig:Data_Processing}(b)] was taken from the fridge before the measurement to induce a more visible temperature drift. This drift can easily be compensated by means of image registration. 


\paragraph*{Movie 3: Real-time binding of 500 pM SAv to [NH-biotin|NH-PEG] molograms.} 
\label{par:movie_3_real_time_binding_of_500_pm_sav_to_nh_biotin_nh_peg_molograms_}

The movie shows the real-time binding of 500 pM SAv in PBS-T pH 7.4 [0.05 \% Tween20] buffer to a [NH-biotin|NH-PEG$_{12}$] mologram. The movie corresponds to the light blue curve in Fig.  \ref{fig8}(b).


\newpage 
\bibliography{references}

\begin{thebibliography}{60}
\expandafter\ifx\csname natexlab\endcsname\relax\def\natexlab#1{#1}\fi
\expandafter\ifx\csname bibnamefont\endcsname\relax
  \def\bibnamefont#1{#1}\fi
\expandafter\ifx\csname bibfnamefont\endcsname\relax
  \def\bibfnamefont#1{#1}\fi
\expandafter\ifx\csname citenamefont\endcsname\relax
  \def\citenamefont#1{#1}\fi
\expandafter\ifx\csname url\endcsname\relax
  \def\url#1{\texttt{#1}}\fi
\expandafter\ifx\csname urlprefix\endcsname\relax\def\urlprefix{URL }\fi
\providecommand{\bibinfo}[2]{#2}
\providecommand{\eprint}[2][]{\url{#2}}

\bibitem[{\citenamefont{Mohammad et~al.}(2018)\citenamefont{Mohammad, Meem,
  Shen, Wang, and Menon}}]{Mohammad2018-bb}
\bibinfo{author}{\bibfnamefont{N.}~\bibnamefont{Mohammad}},
  \bibinfo{author}{\bibfnamefont{M.}~\bibnamefont{Meem}},
  \bibinfo{author}{\bibfnamefont{B.}~\bibnamefont{Shen}},
  \bibinfo{author}{\bibfnamefont{P.}~\bibnamefont{Wang}}, \bibnamefont{and}
  \bibinfo{author}{\bibfnamefont{R.}~\bibnamefont{Menon}},
  \bibinfo{journal}{Sci. Rep.} \textbf{\bibinfo{volume}{8}},
  \bibinfo{pages}{2799} (\bibinfo{year}{2018}).

\bibitem[{\citenamefont{Early et~al.}(2004)\citenamefont{Early, Hyde, and
  Baron}}]{Early2004-gz}
\bibinfo{author}{\bibfnamefont{J.~T.} \bibnamefont{Early}},
  \bibinfo{author}{\bibfnamefont{R.}~\bibnamefont{Hyde}}, \bibnamefont{and}
  \bibinfo{author}{\bibfnamefont{R.~L.} \bibnamefont{Baron}}, in
  \emph{\bibinfo{booktitle}{{UV/Optical/IR} Space Telescopes: Innovative
  Technologies and Concepts}} (\bibinfo{publisher}{International Society for
  Optics and Photonics}, \bibinfo{year}{2004}), vol. \bibinfo{volume}{5166},
  pp. \bibinfo{pages}{148--157}.

\bibitem[{\citenamefont{Park et~al.}(2008)\citenamefont{Park, Koch, Song, Park,
  King, and Choi}}]{Park2008-yn}
\bibinfo{author}{\bibfnamefont{Y.}~\bibnamefont{Park}},
  \bibinfo{author}{\bibfnamefont{L.}~\bibnamefont{Koch}},
  \bibinfo{author}{\bibfnamefont{K.~D.} \bibnamefont{Song}},
  \bibinfo{author}{\bibfnamefont{S.}~\bibnamefont{Park}},
  \bibinfo{author}{\bibfnamefont{G.}~\bibnamefont{King}}, \bibnamefont{and}
  \bibinfo{author}{\bibfnamefont{S.}~\bibnamefont{Choi}}, \bibinfo{journal}{J.
  Opt. A: Pure Appl. Opt.} \textbf{\bibinfo{volume}{10}},
  \bibinfo{pages}{095301} (\bibinfo{year}{2008}).

\bibitem[{\citenamefont{Rodrigues~Ribeiro
  et~al.}(2017)\citenamefont{Rodrigues~Ribeiro, Dahal, Guerreiro, Jorge, and
  Viegas}}]{Rodrigues_Ribeiro2017-jj}
\bibinfo{author}{\bibfnamefont{R.~S.} \bibnamefont{Rodrigues~Ribeiro}},
  \bibinfo{author}{\bibfnamefont{P.}~\bibnamefont{Dahal}},
  \bibinfo{author}{\bibfnamefont{A.}~\bibnamefont{Guerreiro}},
  \bibinfo{author}{\bibfnamefont{P.~A.~S.} \bibnamefont{Jorge}},
  \bibnamefont{and} \bibinfo{author}{\bibfnamefont{J.}~\bibnamefont{Viegas}},
  \bibinfo{journal}{Sci. Rep.} \textbf{\bibinfo{volume}{7}},
  \bibinfo{pages}{4485} (\bibinfo{year}{2017}).

\bibitem[{\citenamefont{Mohacsi et~al.}(2014)\citenamefont{Mohacsi, Karvinen,
  Vartiainen, Guzenko, Somogyi, Kewish, Mercere, and David}}]{Mohacsi2014-ol}
\bibinfo{author}{\bibfnamefont{I.}~\bibnamefont{Mohacsi}},
  \bibinfo{author}{\bibfnamefont{P.}~\bibnamefont{Karvinen}},
  \bibinfo{author}{\bibfnamefont{I.}~\bibnamefont{Vartiainen}},
  \bibinfo{author}{\bibfnamefont{V.~A.} \bibnamefont{Guzenko}},
  \bibinfo{author}{\bibfnamefont{A.}~\bibnamefont{Somogyi}},
  \bibinfo{author}{\bibfnamefont{C.~M.} \bibnamefont{Kewish}},
  \bibinfo{author}{\bibfnamefont{P.}~\bibnamefont{Mercere}}, \bibnamefont{and}
  \bibinfo{author}{\bibfnamefont{C.}~\bibnamefont{David}}, \bibinfo{journal}{J.
  Synchrotron Radiat.} \textbf{\bibinfo{volume}{21}}, \bibinfo{pages}{497}
  (\bibinfo{year}{2014}).

\bibitem[{\citenamefont{Palmer et~al.}(2017)\citenamefont{Palmer, Taylor,
  Brumfeld, Gur, Shemesh, Elad, Osherov, Oron, Weiner, and
  Addadi}}]{Palmer2017-cl}
\bibinfo{author}{\bibfnamefont{B.~A.} \bibnamefont{Palmer}},
  \bibinfo{author}{\bibfnamefont{G.~J.} \bibnamefont{Taylor}},
  \bibinfo{author}{\bibfnamefont{V.}~\bibnamefont{Brumfeld}},
  \bibinfo{author}{\bibfnamefont{D.}~\bibnamefont{Gur}},
  \bibinfo{author}{\bibfnamefont{M.}~\bibnamefont{Shemesh}},
  \bibinfo{author}{\bibfnamefont{N.}~\bibnamefont{Elad}},
  \bibinfo{author}{\bibfnamefont{A.}~\bibnamefont{Osherov}},
  \bibinfo{author}{\bibfnamefont{D.}~\bibnamefont{Oron}},
  \bibinfo{author}{\bibfnamefont{S.}~\bibnamefont{Weiner}}, \bibnamefont{and}
  \bibinfo{author}{\bibfnamefont{L.}~\bibnamefont{Addadi}},
  \bibinfo{journal}{Science} \textbf{\bibinfo{volume}{358}},
  \bibinfo{pages}{1172} (\bibinfo{year}{2017}).

\bibitem[{\citenamefont{Gatterdam et~al.}(2017)\citenamefont{Gatterdam,
  Frutiger, Stengele, Heindl, L{\"u}bbers, V{\"o}r{\"o}s, and
  Fattinger}}]{Gatterdam2017-oo}
\bibinfo{author}{\bibfnamefont{V.}~\bibnamefont{Gatterdam}},
  \bibinfo{author}{\bibfnamefont{A.}~\bibnamefont{Frutiger}},
  \bibinfo{author}{\bibfnamefont{K.-P.} \bibnamefont{Stengele}},
  \bibinfo{author}{\bibfnamefont{D.}~\bibnamefont{Heindl}},
  \bibinfo{author}{\bibfnamefont{T.}~\bibnamefont{L{\"u}bbers}},
  \bibinfo{author}{\bibfnamefont{J.}~\bibnamefont{V{\"o}r{\"o}s}},
  \bibnamefont{and}
  \bibinfo{author}{\bibfnamefont{C.}~\bibnamefont{Fattinger}},
  \bibinfo{journal}{Nat. Nanotechnol.} \textbf{\bibinfo{volume}{12}},
  \bibinfo{pages}{1089} (\bibinfo{year}{2017}).

\bibitem[{\citenamefont{Serrano et~al.}(2016)\citenamefont{Serrano,
  Z{\"u}rcher, Tosatti, and Spencer}}]{Serrano2016-hy}
\bibinfo{author}{\bibfnamefont{{\^A}.}~\bibnamefont{Serrano}},
  \bibinfo{author}{\bibfnamefont{S.}~\bibnamefont{Z{\"u}rcher}},
  \bibinfo{author}{\bibfnamefont{S.}~\bibnamefont{Tosatti}}, \bibnamefont{and}
  \bibinfo{author}{\bibfnamefont{N.~D.} \bibnamefont{Spencer}},
  \bibinfo{journal}{Macromol. Rapid Commun.} \textbf{\bibinfo{volume}{37}},
  \bibinfo{pages}{622} (\bibinfo{year}{2016}).

\bibitem[{\citenamefont{Fraser et~al.}(2017)\citenamefont{Fraser, Shih, Ware,
  O'Connor, Cameron, Schwickart, Zhao, and Regnstrom}}]{Fraser2017-vs}
\bibinfo{author}{\bibfnamefont{S.}~\bibnamefont{Fraser}},
  \bibinfo{author}{\bibfnamefont{J.~Y.} \bibnamefont{Shih}},
  \bibinfo{author}{\bibfnamefont{M.}~\bibnamefont{Ware}},
  \bibinfo{author}{\bibfnamefont{E.}~\bibnamefont{O'Connor}},
  \bibinfo{author}{\bibfnamefont{M.~J.} \bibnamefont{Cameron}},
  \bibinfo{author}{\bibfnamefont{M.}~\bibnamefont{Schwickart}},
  \bibinfo{author}{\bibfnamefont{X.}~\bibnamefont{Zhao}}, \bibnamefont{and}
  \bibinfo{author}{\bibfnamefont{K.}~\bibnamefont{Regnstrom}},
  \bibinfo{journal}{AAPS J.} \textbf{\bibinfo{volume}{19}},
  \bibinfo{pages}{682} (\bibinfo{year}{2017}).

\bibitem[{\citenamefont{{Homola}}(2006)}]{Homola2006-uz}
\bibinfo{author}{\bibnamefont{{Homola}}}, \emph{\bibinfo{title}{Surface Plasmon
  Resonance Based Sensors}}, vol.~\bibinfo{volume}{4} of
  \emph{\bibinfo{series}{Springer Series on Chemical Sensors and Biosensors}}
  (\bibinfo{publisher}{Springer Berlin Heidelberg}, \bibinfo{address}{Berlin,
  Heidelberg}, \bibinfo{year}{2006}).

\bibitem[{\citenamefont{Kozma et~al.}(2014)\citenamefont{Kozma, Kehl,
  Ehrentreich-F{\"o}rster, Stamm, and Bier}}]{Kozma2014-tv}
\bibinfo{author}{\bibfnamefont{P.}~\bibnamefont{Kozma}},
  \bibinfo{author}{\bibfnamefont{F.}~\bibnamefont{Kehl}},
  \bibinfo{author}{\bibfnamefont{E.}~\bibnamefont{Ehrentreich-F{\"o}rster}},
  \bibinfo{author}{\bibfnamefont{C.}~\bibnamefont{Stamm}}, \bibnamefont{and}
  \bibinfo{author}{\bibfnamefont{F.~F.} \bibnamefont{Bier}},
  \bibinfo{journal}{Biosensors and Bioelectronics}
  \textbf{\bibinfo{volume}{58}}, \bibinfo{pages}{287} (\bibinfo{year}{2014}).

\bibitem[{\citenamefont{Cannon et~al.}(2004)\citenamefont{Cannon, Papalia,
  Navratilova, Fisher, Roberts, Worthy, Stephen, Marchesini, Collins, Casper
  et~al.}}]{Cannon2004-kf}
\bibinfo{author}{\bibfnamefont{M.~J.} \bibnamefont{Cannon}},
  \bibinfo{author}{\bibfnamefont{G.~A.} \bibnamefont{Papalia}},
  \bibinfo{author}{\bibfnamefont{I.}~\bibnamefont{Navratilova}},
  \bibinfo{author}{\bibfnamefont{R.~J.} \bibnamefont{Fisher}},
  \bibinfo{author}{\bibfnamefont{L.~R.} \bibnamefont{Roberts}},
  \bibinfo{author}{\bibfnamefont{K.~M.} \bibnamefont{Worthy}},
  \bibinfo{author}{\bibfnamefont{A.~G.} \bibnamefont{Stephen}},
  \bibinfo{author}{\bibfnamefont{G.~R.} \bibnamefont{Marchesini}},
  \bibinfo{author}{\bibfnamefont{E.~J.} \bibnamefont{Collins}},
  \bibinfo{author}{\bibfnamefont{D.}~\bibnamefont{Casper}},
  \bibnamefont{et~al.}, \bibinfo{journal}{Anal. Biochem.}
  \textbf{\bibinfo{volume}{330}}, \bibinfo{pages}{98} (\bibinfo{year}{2004}).

\bibitem[{\citenamefont{Kozma et~al.}(2009)\citenamefont{Kozma, Hamori,
  Cottier, Kurunczi, and Horvath}}]{Kozma2009-tr}
\bibinfo{author}{\bibfnamefont{P.}~\bibnamefont{Kozma}},
  \bibinfo{author}{\bibfnamefont{A.}~\bibnamefont{Hamori}},
  \bibinfo{author}{\bibfnamefont{K.}~\bibnamefont{Cottier}},
  \bibinfo{author}{\bibfnamefont{S.}~\bibnamefont{Kurunczi}}, \bibnamefont{and}
  \bibinfo{author}{\bibfnamefont{R.}~\bibnamefont{Horvath}},
  \bibinfo{journal}{Appl. Phys. B} \textbf{\bibinfo{volume}{97}},
  \bibinfo{pages}{5} (\bibinfo{year}{2009}).

\bibitem[{\citenamefont{Piliarik et~al.}(2010)\citenamefont{Piliarik,
  Bockov{\'a}, and Homola}}]{Piliarik2010-fb}
\bibinfo{author}{\bibfnamefont{M.}~\bibnamefont{Piliarik}},
  \bibinfo{author}{\bibfnamefont{M.}~\bibnamefont{Bockov{\'a}}},
  \bibnamefont{and} \bibinfo{author}{\bibfnamefont{J.}~\bibnamefont{Homola}},
  \bibinfo{journal}{Biosens. Bioelectron.} \textbf{\bibinfo{volume}{26}},
  \bibinfo{pages}{1656} (\bibinfo{year}{2010}).

\bibitem[{\citenamefont{Piliarik and Sandoghdar}(2014)}]{Piliarik2014-ys}
\bibinfo{author}{\bibfnamefont{M.}~\bibnamefont{Piliarik}} \bibnamefont{and}
  \bibinfo{author}{\bibfnamefont{V.}~\bibnamefont{Sandoghdar}},
  \bibinfo{journal}{Nat. Commun.} \textbf{\bibinfo{volume}{5}},
  \bibinfo{pages}{4495} (\bibinfo{year}{2014}).

\bibitem[{\citenamefont{Fattinger}(2014)}]{Fattinger2014-dm}
\bibinfo{author}{\bibfnamefont{C.}~\bibnamefont{Fattinger}},
  \bibinfo{journal}{Phys. Rev. X} \textbf{\bibinfo{volume}{4}},
  \bibinfo{pages}{031024} (\bibinfo{year}{2014}).

\bibitem[{\citenamefont{Pasche et~al.}(2003)\citenamefont{Pasche, De~Paul,
  Voros, Spencer, and Textor}}]{Pasche2003-mr}
\bibinfo{author}{\bibfnamefont{S.}~\bibnamefont{Pasche}},
  \bibinfo{author}{\bibfnamefont{S.~M.} \bibnamefont{De~Paul}},
  \bibinfo{author}{\bibfnamefont{J.}~\bibnamefont{Voros}},
  \bibinfo{author}{\bibfnamefont{N.~D.} \bibnamefont{Spencer}},
  \bibnamefont{and} \bibinfo{author}{\bibfnamefont{M.}~\bibnamefont{Textor}},
  \bibinfo{journal}{Langmuir} \textbf{\bibinfo{volume}{19}},
  \bibinfo{pages}{9216} (\bibinfo{year}{2003}).

\bibitem[{\citenamefont{Love et~al.}(2005)\citenamefont{Love, Estroff, Kriebel,
  Nuzzo, and Whitesides}}]{Love2005-oq}
\bibinfo{author}{\bibfnamefont{J.~C.} \bibnamefont{Love}},
  \bibinfo{author}{\bibfnamefont{L.~A.} \bibnamefont{Estroff}},
  \bibinfo{author}{\bibfnamefont{J.~K.} \bibnamefont{Kriebel}},
  \bibinfo{author}{\bibfnamefont{R.~G.} \bibnamefont{Nuzzo}}, \bibnamefont{and}
  \bibinfo{author}{\bibfnamefont{G.~M.} \bibnamefont{Whitesides}},
  \bibinfo{journal}{Chem. Rev.} \textbf{\bibinfo{volume}{105}},
  \bibinfo{pages}{1103} (\bibinfo{year}{2005}).

\bibitem[{\citenamefont{Falconnet et~al.}(2004)\citenamefont{Falconnet, Koenig,
  Assi, and Textor}}]{Falconnet2004-mq}
\bibinfo{author}{\bibfnamefont{D.}~\bibnamefont{Falconnet}},
  \bibinfo{author}{\bibfnamefont{A.}~\bibnamefont{Koenig}},
  \bibinfo{author}{\bibfnamefont{F.}~\bibnamefont{Assi}}, \bibnamefont{and}
  \bibinfo{author}{\bibfnamefont{M.}~\bibnamefont{Textor}},
  \bibinfo{journal}{Adv. Funct. Mater.} \textbf{\bibinfo{volume}{14}},
  \bibinfo{pages}{749} (\bibinfo{year}{2004}).

\bibitem[{\citenamefont{Vigneswaran et~al.}(2014)\citenamefont{Vigneswaran,
  Samsuri, Ranganathan, and {Padmapriya}}}]{Vigneswaran2014-ka}
\bibinfo{author}{\bibfnamefont{N.}~\bibnamefont{Vigneswaran}},
  \bibinfo{author}{\bibfnamefont{F.}~\bibnamefont{Samsuri}},
  \bibinfo{author}{\bibfnamefont{B.}~\bibnamefont{Ranganathan}},
  \bibnamefont{and} \bibinfo{author}{\bibnamefont{{Padmapriya}}},
  \bibinfo{journal}{Procedia Engineering} \textbf{\bibinfo{volume}{97}},
  \bibinfo{pages}{1387} (\bibinfo{year}{2014}).

\bibitem[{\citenamefont{Avella-Oliver et~al.}(2017)\citenamefont{Avella-Oliver,
  Carrascosa, Puchades, and Maquieira}}]{Avella-Oliver2017-jb}
\bibinfo{author}{\bibfnamefont{M.}~\bibnamefont{Avella-Oliver}},
  \bibinfo{author}{\bibfnamefont{J.}~\bibnamefont{Carrascosa}},
  \bibinfo{author}{\bibfnamefont{R.}~\bibnamefont{Puchades}}, \bibnamefont{and}
  \bibinfo{author}{\bibfnamefont{{\'A}.}~\bibnamefont{Maquieira}},
  \bibinfo{journal}{Anal. Chem.} \textbf{\bibinfo{volume}{89}},
  \bibinfo{pages}{9002} (\bibinfo{year}{2017}).

\bibitem[{\citenamefont{Cleverley et~al.}(2010)\citenamefont{Cleverley, Chen,
  and Houle}}]{Cleverley2010-qm}
\bibinfo{author}{\bibfnamefont{S.}~\bibnamefont{Cleverley}},
  \bibinfo{author}{\bibfnamefont{I.}~\bibnamefont{Chen}}, \bibnamefont{and}
  \bibinfo{author}{\bibfnamefont{J.-F.} \bibnamefont{Houle}},
  \bibinfo{journal}{J. Chromatogr. B Analyt. Technol. Biomed. Life Sci.}
  \textbf{\bibinfo{volume}{878}}, \bibinfo{pages}{264} (\bibinfo{year}{2010}).

\bibitem[{\citenamefont{Lenhert et~al.}(2010)\citenamefont{Lenhert, Brinkmann,
  Laue, Walheim, Vannahme, Klinkhammer, Xu, Sekula, Mappes, Schimmel
  et~al.}}]{Lenhert2010-tt}
\bibinfo{author}{\bibfnamefont{S.}~\bibnamefont{Lenhert}},
  \bibinfo{author}{\bibfnamefont{F.}~\bibnamefont{Brinkmann}},
  \bibinfo{author}{\bibfnamefont{T.}~\bibnamefont{Laue}},
  \bibinfo{author}{\bibfnamefont{S.}~\bibnamefont{Walheim}},
  \bibinfo{author}{\bibfnamefont{C.}~\bibnamefont{Vannahme}},
  \bibinfo{author}{\bibfnamefont{S.}~\bibnamefont{Klinkhammer}},
  \bibinfo{author}{\bibfnamefont{M.}~\bibnamefont{Xu}},
  \bibinfo{author}{\bibfnamefont{S.}~\bibnamefont{Sekula}},
  \bibinfo{author}{\bibfnamefont{T.}~\bibnamefont{Mappes}},
  \bibinfo{author}{\bibfnamefont{T.}~\bibnamefont{Schimmel}},
  \bibnamefont{et~al.}, \bibinfo{journal}{Nat. Nanotechnol.}
  \textbf{\bibinfo{volume}{5}}, \bibinfo{pages}{275} (\bibinfo{year}{2010}).

\bibitem[{\citenamefont{Lai et~al.}(2008)\citenamefont{Lai, Wang, Allbritton,
  Li, and Bachman}}]{Lai2008-ha}
\bibinfo{author}{\bibfnamefont{Z.}~\bibnamefont{Lai}},
  \bibinfo{author}{\bibfnamefont{Y.}~\bibnamefont{Wang}},
  \bibinfo{author}{\bibfnamefont{N.}~\bibnamefont{Allbritton}},
  \bibinfo{author}{\bibfnamefont{G.-P.} \bibnamefont{Li}}, \bibnamefont{and}
  \bibinfo{author}{\bibfnamefont{M.}~\bibnamefont{Bachman}},
  \bibinfo{journal}{Opt. Lett.} \textbf{\bibinfo{volume}{33}},
  \bibinfo{pages}{1735} (\bibinfo{year}{2008}).

\bibitem[{\citenamefont{Kogelnik}(1969)}]{Kogelnik1969-pf}
\bibinfo{author}{\bibfnamefont{H.}~\bibnamefont{Kogelnik}},
  \bibinfo{journal}{The Bell System Technical Journal}
  \textbf{\bibinfo{volume}{48}} (\bibinfo{year}{1969}).

\bibitem[{\citenamefont{Poole}(1998)}]{Poole1998-pk}
\bibinfo{author}{\bibfnamefont{I.}~\bibnamefont{Poole}},
  \emph{\bibinfo{title}{Basic Radio: Principles and Technology}}
  (\bibinfo{publisher}{Newnes}, \bibinfo{year}{1998}).

\bibitem[{noa()}]{noauthor_undated-ex}
\emph{\bibinfo{title}{Selection guide biacore systems}},
  \bibinfo{howpublished}{\url{https://proteins.gelifesciences.com/}},
  \bibinfo{note}{accessed: 2017-12-1}.

\bibitem[{\citenamefont{Marcuse}(1974)}]{Marcuse1974-uo}
\bibinfo{author}{\bibfnamefont{D.}~\bibnamefont{Marcuse}},
  \emph{\bibinfo{title}{Theory of Dielectric Optical Waveguides}}
  (\bibinfo{publisher}{Academic Press}, \bibinfo{year}{1974}).

\bibitem[{\citenamefont{Novotny and Hecht}(2011)}]{Novotny2011-pa}
\bibinfo{author}{\bibfnamefont{L.}~\bibnamefont{Novotny}} \bibnamefont{and}
  \bibinfo{author}{\bibfnamefont{B.}~\bibnamefont{Hecht}},
  \emph{\bibinfo{title}{Principles of {Nano-Optics}}}
  (\bibinfo{publisher}{Cambridge University Press}, \bibinfo{year}{2011}).

\bibitem[{\citenamefont{Goodman}(2005)}]{Goodman2005-wy}
\bibinfo{author}{\bibfnamefont{J.~W.} \bibnamefont{Goodman}},
  \emph{\bibinfo{title}{Introduction to Fourier Optics}}
  (\bibinfo{publisher}{Roberts and Company Publishers}, \bibinfo{year}{2005}).

\bibitem[{\citenamefont{Sinclair}(1947)}]{Sinclair1947-xa}
\bibinfo{author}{\bibfnamefont{D.}~\bibnamefont{Sinclair}},
  \bibinfo{journal}{J. Opt. Soc. Am.} \textbf{\bibinfo{volume}{37}},
  \bibinfo{pages}{475} (\bibinfo{year}{1947}).

\bibitem[{\citenamefont{Monneret et~al.}(2000)\citenamefont{Monneret,
  Huguet-Chant{\^o}me, and Flory}}]{Monneret2000-im}
\bibinfo{author}{\bibfnamefont{S.}~\bibnamefont{Monneret}},
  \bibinfo{author}{\bibfnamefont{P.}~\bibnamefont{Huguet-Chant{\^o}me}},
  \bibnamefont{and} \bibinfo{author}{\bibfnamefont{F.}~\bibnamefont{Flory}},
  \bibinfo{journal}{J. Opt. A: Pure Appl. Opt.} \textbf{\bibinfo{volume}{2}},
  \bibinfo{pages}{188} (\bibinfo{year}{2000}).

\bibitem[{\citenamefont{Tamir and Peng}(1977)}]{Tamir1977-xc}
\bibinfo{author}{\bibfnamefont{T.}~\bibnamefont{Tamir}} \bibnamefont{and}
  \bibinfo{author}{\bibfnamefont{S.~T.} \bibnamefont{Peng}},
  \bibinfo{journal}{J. Phys. D Appl. Phys.} \textbf{\bibinfo{volume}{14}},
  \bibinfo{pages}{235} (\bibinfo{year}{1977}).

\bibitem[{\citenamefont{Zhao et~al.}(2011)\citenamefont{Zhao, Brown, and
  Schuck}}]{Zhao2011-ju}
\bibinfo{author}{\bibfnamefont{H.}~\bibnamefont{Zhao}},
  \bibinfo{author}{\bibfnamefont{P.~H.} \bibnamefont{Brown}}, \bibnamefont{and}
  \bibinfo{author}{\bibfnamefont{P.}~\bibnamefont{Schuck}},
  \bibinfo{journal}{Biophys. J.} \textbf{\bibinfo{volume}{100}},
  \bibinfo{pages}{2309} (\bibinfo{year}{2011}).

\bibitem[{\citenamefont{Magnusson and Gaylord}(1978)}]{Magnusson1978-sx}
\bibinfo{author}{\bibfnamefont{R.}~\bibnamefont{Magnusson}} \bibnamefont{and}
  \bibinfo{author}{\bibfnamefont{T.~K.} \bibnamefont{Gaylord}},
  \bibinfo{journal}{J. Opt. Soc. Am., JOSA} \textbf{\bibinfo{volume}{68}},
  \bibinfo{pages}{806} (\bibinfo{year}{1978}).

\bibitem[{\citenamefont{Heller}(1965)}]{Heller1965-me}
\bibinfo{author}{\bibfnamefont{W.}~\bibnamefont{Heller}}, \bibinfo{journal}{J.
  Phys. Chem.} \textbf{\bibinfo{volume}{69}}, \bibinfo{pages}{1123}
  (\bibinfo{year}{1965}).

\bibitem[{\citenamefont{Fischer et~al.}(2004)\citenamefont{Fischer, Polikarpov,
  and Craievich}}]{Fischer2004-io}
\bibinfo{author}{\bibfnamefont{H.}~\bibnamefont{Fischer}},
  \bibinfo{author}{\bibfnamefont{I.}~\bibnamefont{Polikarpov}},
  \bibnamefont{and} \bibinfo{author}{\bibfnamefont{A.~F.}
  \bibnamefont{Craievich}}, \bibinfo{journal}{Protein Sci.}
  \textbf{\bibinfo{volume}{13}}, \bibinfo{pages}{2825} (\bibinfo{year}{2004}).

\bibitem[{\citenamefont{Lukosz}(1981)}]{Lukosz1981-js}
\bibinfo{author}{\bibfnamefont{W.}~\bibnamefont{Lukosz}}, \bibinfo{journal}{J.
  Opt. Soc. Am., JOSA} \textbf{\bibinfo{volume}{71}}, \bibinfo{pages}{744}
  (\bibinfo{year}{1981}).

\bibitem[{\citenamefont{{Structural Genomics Consortium}
  et~al.}(2008)\citenamefont{{Structural Genomics Consortium}, {China
  Structural Genomics Consortium}, {Northeast Structural Genomics Consortium},
  Gr{\"a}slund, Nordlund, Weigelt, Hallberg, Bray, Gileadi, Knapp
  et~al.}}]{Structural_Genomics_Consortium2008-hk}
\bibinfo{author}{\bibnamefont{{Structural Genomics Consortium}}},
  \bibinfo{author}{\bibnamefont{{China Structural Genomics Consortium}}},
  \bibinfo{author}{\bibnamefont{{Northeast Structural Genomics Consortium}}},
  \bibinfo{author}{\bibfnamefont{S.}~\bibnamefont{Gr{\"a}slund}},
  \bibinfo{author}{\bibfnamefont{P.}~\bibnamefont{Nordlund}},
  \bibinfo{author}{\bibfnamefont{J.}~\bibnamefont{Weigelt}},
  \bibinfo{author}{\bibfnamefont{B.~M.} \bibnamefont{Hallberg}},
  \bibinfo{author}{\bibfnamefont{J.}~\bibnamefont{Bray}},
  \bibinfo{author}{\bibfnamefont{O.}~\bibnamefont{Gileadi}},
  \bibinfo{author}{\bibfnamefont{S.}~\bibnamefont{Knapp}},
  \bibnamefont{et~al.}, \bibinfo{journal}{Nat. Methods}
  \textbf{\bibinfo{volume}{5}}, \bibinfo{pages}{135} (\bibinfo{year}{2008}).

\bibitem[{\citenamefont{Dainty}(1977)}]{Dainty1977-nh}
\bibinfo{author}{\bibfnamefont{J.~C.} \bibnamefont{Dainty}}, in
  \emph{\bibinfo{booktitle}{Progress in Optics}}, edited by
  \bibinfo{editor}{\bibfnamefont{E.}~\bibnamefont{Wolf}}
  (\bibinfo{publisher}{Elsevier}, \bibinfo{year}{1977}),
  vol.~\bibinfo{volume}{14}, pp. \bibinfo{pages}{1--46}.

\bibitem[{\citenamefont{Miyanaga et~al.}(1980)\citenamefont{Miyanaga, Asakura,
  and Imai}}]{Miyanaga1980-od}
\bibinfo{author}{\bibfnamefont{S.}~\bibnamefont{Miyanaga}},
  \bibinfo{author}{\bibfnamefont{T.}~\bibnamefont{Asakura}}, \bibnamefont{and}
  \bibinfo{author}{\bibfnamefont{M.}~\bibnamefont{Imai}},
  \bibinfo{journal}{Opt. Quantum Electron.} \textbf{\bibinfo{volume}{12}},
  \bibinfo{pages}{23} (\bibinfo{year}{1980}).

\bibitem[{\citenamefont{Zhang et~al.}(2007)\citenamefont{Zhang, Dammann, Bae,
  and Granick}}]{Zhang2007-bb}
\bibinfo{author}{\bibfnamefont{L.}~\bibnamefont{Zhang}},
  \bibinfo{author}{\bibfnamefont{K.}~\bibnamefont{Dammann}},
  \bibinfo{author}{\bibfnamefont{S.~C.} \bibnamefont{Bae}}, \bibnamefont{and}
  \bibinfo{author}{\bibfnamefont{S.}~\bibnamefont{Granick}},
  \bibinfo{journal}{Soft Matter} \textbf{\bibinfo{volume}{3}},
  \bibinfo{pages}{551} (\bibinfo{year}{2007}).

\bibitem[{\citenamefont{Liu et~al.}(2012)\citenamefont{Liu, Ji, Tao, Tan,
  Zhang, and Fu}}]{Liu2012-sx}
\bibinfo{author}{\bibfnamefont{S.-Q.} \bibnamefont{Liu}},
  \bibinfo{author}{\bibfnamefont{X.-L.} \bibnamefont{Ji}},
  \bibinfo{author}{\bibfnamefont{Y.}~\bibnamefont{Tao}},
  \bibinfo{author}{\bibfnamefont{D.-Y.} \bibnamefont{Tan}},
  \bibinfo{author}{\bibfnamefont{K.-Q.} \bibnamefont{Zhang}}, \bibnamefont{and}
  \bibinfo{author}{\bibfnamefont{Y.-X.} \bibnamefont{Fu}}, in
  \emph{\bibinfo{booktitle}{Protein engineering}} (\bibinfo{publisher}{InTech},
  \bibinfo{year}{2012}).

\bibitem[{\citenamefont{Hahn}(2009)}]{Hahn2009-sp}
\bibinfo{author}{\bibfnamefont{D.~W.} \bibnamefont{Hahn}},
  \bibinfo{journal}{Department of Mechanical and Aerospace Engineering,
  University of Florida}  (\bibinfo{year}{2009}).

\bibitem[{\citenamefont{Ladouceur and Love}(1996)}]{Ladouceur1996-of}
\bibinfo{author}{\bibfnamefont{F.}~\bibnamefont{Ladouceur}} \bibnamefont{and}
  \bibinfo{author}{\bibfnamefont{J.~D.} \bibnamefont{Love}},
  \emph{\bibinfo{title}{Silica-based buried channel waveguides and devices}}
  (\bibinfo{publisher}{Chapman \& Hall}, \bibinfo{year}{1996}).

\bibitem[{\citenamefont{Payne and Lacey}(1994)}]{Payne1994-tw}
\bibinfo{author}{\bibfnamefont{F.~P.} \bibnamefont{Payne}} \bibnamefont{and}
  \bibinfo{author}{\bibfnamefont{J.~P.~R.} \bibnamefont{Lacey}},
  \bibinfo{journal}{Opt. Quantum Electron.} \textbf{\bibinfo{volume}{26}},
  \bibinfo{pages}{977} (\bibinfo{year}{1994}).

\bibitem[{\citenamefont{Ames and Hall}(1983)}]{Ames1983-bn}
\bibinfo{author}{\bibfnamefont{G.}~\bibnamefont{Ames}} \bibnamefont{and}
  \bibinfo{author}{\bibfnamefont{D.}~\bibnamefont{Hall}},
  \bibinfo{journal}{IEEE J. Quantum Electron.} \textbf{\bibinfo{volume}{19}},
  \bibinfo{pages}{845} (\bibinfo{year}{1983}).

\bibitem[{\citenamefont{Miyanaga et~al.}(1979)\citenamefont{Miyanaga, Asakura,
  and Imai}}]{Miyanaga1979-ht}
\bibinfo{author}{\bibfnamefont{S.}~\bibnamefont{Miyanaga}},
  \bibinfo{author}{\bibfnamefont{T.}~\bibnamefont{Asakura}}, \bibnamefont{and}
  \bibinfo{author}{\bibfnamefont{M.}~\bibnamefont{Imai}},
  \bibinfo{journal}{Opt. Quantum Electron.} \textbf{\bibinfo{volume}{11}},
  \bibinfo{pages}{205} (\bibinfo{year}{1979}).

\bibitem[{\citenamefont{Ronald~Louis}(1992)}]{Ronald_Louis1992-qj}
\bibinfo{author}{\bibfnamefont{R.}~\bibnamefont{Ronald~Louis}}, Ph.D. thesis,
  \bibinfo{school}{The University of Arizona} (\bibinfo{year}{1992}).

\bibitem[{\citenamefont{Lacey and Payne}(1990)}]{Lacey1990-ii}
\bibinfo{author}{\bibfnamefont{J.}~\bibnamefont{Lacey}} \bibnamefont{and}
  \bibinfo{author}{\bibfnamefont{F.~P.} \bibnamefont{Payne}},
  \bibinfo{journal}{IEE Proceedings J-Optoelectronics}  (\bibinfo{year}{1990}).

\bibitem[{\citenamefont{Armbruster and Pry}(2008)}]{Armbruster2008-ol}
\bibinfo{author}{\bibfnamefont{D.~A.} \bibnamefont{Armbruster}}
  \bibnamefont{and} \bibinfo{author}{\bibfnamefont{T.}~\bibnamefont{Pry}},
  \bibinfo{journal}{Clin. Biochem. Rev.} \textbf{\bibinfo{volume}{29 Suppl 1}},
  \bibinfo{pages}{S49} (\bibinfo{year}{2008}).

\bibitem[{\citenamefont{Holstein et~al.}(2015)\citenamefont{Holstein, Griffin,
  Hong, and Sampson}}]{Holstein2015-ej}
\bibinfo{author}{\bibfnamefont{C.~A.} \bibnamefont{Holstein}},
  \bibinfo{author}{\bibfnamefont{M.}~\bibnamefont{Griffin}},
  \bibinfo{author}{\bibfnamefont{J.}~\bibnamefont{Hong}}, \bibnamefont{and}
  \bibinfo{author}{\bibfnamefont{P.~D.} \bibnamefont{Sampson}},
  \bibinfo{journal}{Anal. Chem.} \textbf{\bibinfo{volume}{87}},
  \bibinfo{pages}{9795} (\bibinfo{year}{2015}).

\bibitem[{\citenamefont{Homola}(2008)}]{Homola2008-px}
\bibinfo{author}{\bibfnamefont{J.}~\bibnamefont{Homola}},
  \bibinfo{journal}{Chem. Rev.} \textbf{\bibinfo{volume}{108}},
  \bibinfo{pages}{462} (\bibinfo{year}{2008}).

\bibitem[{\citenamefont{Piliarik and Homola}(2009)}]{Piliarik2009-xb}
\bibinfo{author}{\bibfnamefont{M.}~\bibnamefont{Piliarik}} \bibnamefont{and}
  \bibinfo{author}{\bibfnamefont{J.}~\bibnamefont{Homola}},
  \bibinfo{journal}{Opt. Express} \textbf{\bibinfo{volume}{17}},
  \bibinfo{pages}{16505} (\bibinfo{year}{2009}).

\bibitem[{\citenamefont{Schasfoort}(2017)}]{Schasfoort2017-hs}
\bibinfo{editor}{\bibfnamefont{R.~B.~M.} \bibnamefont{Schasfoort}}, ed.,
  \emph{\bibinfo{title}{Handbook of Surface Plasmon Resonance}}
  (\bibinfo{publisher}{The Royal Society of Chemistry}, \bibinfo{year}{2017}).

\bibitem[{\citenamefont{Kolomenskii et~al.}(1997)\citenamefont{Kolomenskii,
  Gershon, and Schuessler}}]{Kolomenskii1997-go}
\bibinfo{author}{\bibfnamefont{A.~A.} \bibnamefont{Kolomenskii}},
  \bibinfo{author}{\bibfnamefont{P.~D.} \bibnamefont{Gershon}},
  \bibnamefont{and} \bibinfo{author}{\bibfnamefont{H.~A.}
  \bibnamefont{Schuessler}}, \bibinfo{journal}{Appl. Opt.}
  \textbf{\bibinfo{volume}{36}}, \bibinfo{pages}{6539} (\bibinfo{year}{1997}).

\bibitem[{\citenamefont{Thevenaz et~al.}(1998)\citenamefont{Thevenaz,
  Ruttimann, and Unser}}]{Thevenaz1998-lb}
\bibinfo{author}{\bibfnamefont{P.}~\bibnamefont{Thevenaz}},
  \bibinfo{author}{\bibfnamefont{U.~E.} \bibnamefont{Ruttimann}},
  \bibnamefont{and} \bibinfo{author}{\bibfnamefont{M.}~\bibnamefont{Unser}},
  \bibinfo{journal}{IEEE Trans. Image Process.} \textbf{\bibinfo{volume}{7}},
  \bibinfo{pages}{27} (\bibinfo{year}{1998}).

\bibitem[{\citenamefont{Squires et~al.}(2008)\citenamefont{Squires, Messinger,
  and Manalis}}]{Squires2008-pt}
\bibinfo{author}{\bibfnamefont{T.~M.} \bibnamefont{Squires}},
  \bibinfo{author}{\bibfnamefont{R.~J.} \bibnamefont{Messinger}},
  \bibnamefont{and} \bibinfo{author}{\bibfnamefont{S.~R.}
  \bibnamefont{Manalis}}, \bibinfo{journal}{Nat. Biotechnol.}
  \textbf{\bibinfo{volume}{26}}, \bibinfo{pages}{417} (\bibinfo{year}{2008}).

\bibitem[{\citenamefont{Surinova et~al.}(2011)\citenamefont{Surinova, Schiess,
  H{\"u}ttenhain, Cerciello, Wollscheid, and Aebersold}}]{Surinova2011-uf}
\bibinfo{author}{\bibfnamefont{S.}~\bibnamefont{Surinova}},
  \bibinfo{author}{\bibfnamefont{R.}~\bibnamefont{Schiess}},
  \bibinfo{author}{\bibfnamefont{R.}~\bibnamefont{H{\"u}ttenhain}},
  \bibinfo{author}{\bibfnamefont{F.}~\bibnamefont{Cerciello}},
  \bibinfo{author}{\bibfnamefont{B.}~\bibnamefont{Wollscheid}},
  \bibnamefont{and}
  \bibinfo{author}{\bibfnamefont{R.}~\bibnamefont{Aebersold}},
  \bibinfo{journal}{J. Proteome Res.} \textbf{\bibinfo{volume}{10}},
  \bibinfo{pages}{5} (\bibinfo{year}{2011}).

\bibitem[{\citenamefont{Landry et~al.}(2015)\citenamefont{Landry, Ke, Yu, and
  Zhu}}]{Landry2015-xe}
\bibinfo{author}{\bibfnamefont{J.~P.} \bibnamefont{Landry}},
  \bibinfo{author}{\bibfnamefont{Y.}~\bibnamefont{Ke}},
  \bibinfo{author}{\bibfnamefont{G.-L.} \bibnamefont{Yu}}, \bibnamefont{and}
  \bibinfo{author}{\bibfnamefont{X.~D.} \bibnamefont{Zhu}},
  \bibinfo{journal}{J. Immunol. Methods} \textbf{\bibinfo{volume}{417}},
  \bibinfo{pages}{86} (\bibinfo{year}{2015}).

\end{thebibliography}

\end{document}